\documentclass[a4paper,11pt]{article}
\pdfoutput=1 
\usepackage{jheppub} 
\usepackage[T1]{fontenc} 
\usepackage{graphicx}
\usepackage{placeins}
\usepackage{xspace}
\usepackage{soul}
\usepackage{siunitx}
\usepackage{upgreek}
\usepackage{subcaption}
\usepackage[dvipsnames]{xcolor}
\usepackage{slashed}

\def\mzp{\ifmmode m_{Z^\prime} \else $m_{Z^{\prime}}$ \fi}
\def\gp{\ifmmode g^{\prime} \else $g^{\prime}$ \fi}

\preprint{\begin{flushright} 
\end{flushright}}

\title{\boldmath 
Leptophilic $Z'$ bosons at the FCC-ee: discovery opportunities
}
\author[a]{Rebeca Gonzalez Suarez,} 
\author[b]{Baibhab Pattnaik,}
\author[b]{Jos\'e Zurita}

\affiliation[a]{Department of Physics and Astronomy, Division of High Energy Physics, Uppsala University}
\affiliation[b]{Instituto de F\'{\i}sica Corpuscular, CSIC-Universitat de Val\`encia, \\ Catedr\'{a}tico Jos\'{e} Beltr\'{a}n 2, E-46980, Valencia, Spain}

\emailAdd{rebeca.gonzalez.suarez@cern.ch}
\emailAdd{Baibhab.Pattnaik@ific.uv.es}
\emailAdd{Jose.Zurita@ific.uv.es}

\abstract{We examine the possibility to detect new SM-neutral vector bosons ($Z'$) that couple exclusively to leptons in the electron-positron mode of the Future Circular Collider (FCC-ee). Focusing on the $Z'$ production with a radiated photon search channel, we show that the FCC-ee can significantly extend the unprobed parameter space by increasing the exclusion in the coupling by one to two orders of magnitude in the kinematically allowed mass range (from 10 GeV to 365 GeV), with the leading sensitivity being driven by the muon channel. In doing so, it outperforms other proposed lepton collider options such as CLIC and ILC in this range of masses. Further, we discuss the possibility of improving the sensitivity of the FCC-ee to this model through the modification of the dilepton invariant mass resolution and the photon energy resolution. The impact of systematic uncertainties on the expected sensitivities is also studied. 

}
\begin{document} 
\maketitle
\flushbottom

\section{Introduction}
\label{s.intro}
The possibility that new physics couples to the Standard Model (SM) exclusively via leptons is a phenomenologically appealing scenario. The existing LHC bounds are rather mild in that case, hence providing an important physics motivation for the next generation of lepton colliders, such as the FCC-ee \cite{FCC:2018evy}, ILC \cite{ILC:2013jhg}, CEPC \cite{CEPCStudyGroup:2018ghi}, CLIC \cite{CLICdp:2018cto}, and Muon Collider~\cite{MuonCollider:2022xlm} proposals.

An economical setup for new physics (leptophilic or otherwise) is achieved via the addition of a new vector boson of an Abelian $U(1)$ group. While the gauge charge assignments are, in principle, free if one does not impose unitarity, the cancellation of gauge anomalies is a desirable theoretical ingredient. This leaves only a select few possibilities to choose a SM subgroup to gauge if one wants to keep only the SM matter content (and potentially add right handed neutrinos). The $B-$L~\cite{Mohapatra:1980de} models and $L_x - L_y$~\cite{Foot:1990mn,He:1990pn,Foot:1994vd} models, with $x,y$ being two different lepton families\footnote{And of course linear combinations of the models above; note that $L_e - L_{\tau} = L_e - L_{\mu} + L_{\mu} - L_{\tau} $, hence there are only two independent groups to gauge.} that satisfy these criteria. It may be noted that these models are not just an artifact of minimality, resting only on the automatic gauge anomaly cancellation. Rather, they are well-motivated in that they solve concrete, open problems of the Standard Model of particle physics such as the neutrino mass hierarchies and mixing patterns, provide a portal to dark matter, and even accommodate excesses in the $(g-2)_{\mu}$ \cite{Biswas:2016yan,Baek:2015mna,Heeck:2011wj,Patra:2016shz,Arcadi:2018tly,Altmannshofer:2016jzy,Gninenko:2001hx,Ma:2001md,Baek:2001kca}. Previous work has explored the phenomenological consequences of this setup at lepton colliders running at energy scales of 3 TeV~\cite{Dasgupta:2023zrh,Barik:2024kwv}. In this paper we explore the complementarity obtained with FCC-ee, that runs at a relatively modest center-of-mass energy of up to 365 GeV.\footnote{The CEPC proposal would also run at similar energies. For a recent study on the expected bounds on these models from $e^+ e^- \to l^+ l^- Z'$ see~\cite{Yue:2024kwo}.} We will consider here $L_e - L_{\mu}$ and $L_e - L_{\tau}$ models, which automatically conserve lepton flavour.\footnote{Recent work has explored the FCC-ee prospects for a lepton-flavour violating setup~\cite{Goudelis:2023yni}. 
}. 

The leading signals of this setup at the FCC-ee will arise from the associated production of the $Z'$ with an initial state radiated (ISR) photon, as it sets the $Z'$ on-shell. Lepton pair production is off-shell and hence subdominant unless the $Z'$ mass is very close to one of the planned center-of-mass energies, $1 - s/ \mzp^2 \lesssim 10^{-4} $ considering current constraints. In what follows,  we analyze in detail the FCC-ee sensitivities to the aforementioned lepton-flavour conserving decays, confronting those prospects with the existing and projected constraints on the coupling vs. mass parameter space.  In particular, we show that the FCC-ee probes into unconstrained parameter space throughout the mass range allowed by its center-of-mass energy, i.e. up to $Z'$ masses of 365 GeV. Additionally, we examine how parameters such as the di-lepton mass resolution and photon energy resolution affect the final sensitivities, providing crucial input for the FCC-ee detector capabilities.
This article is structured as follows.  In section~\ref{s.theory} we review the basic phenomenological features of the $Z'$ scenarios under consideration, and review the existing (and future) constraints on the parameter space. In section~\ref{s.coll} we present our studies for the FCC-ee, focusing on  $Z'$ production with a radiated photon. Our conclusions are presented in section~\ref{s.conclu}.

\section{Leptophilic $Z'$ models}
\label{s.theory}
The Lagrangian for a leptophilic $Z'$ with gauge group $U(1)_{L_x - L_y}$ and vector couplings to the SM leptons reads 
\begin{eqnarray}
\label{e.lagzpl}
{\cal L}
& \supset & - \gp  \left(\bar{l}_x  \slashed{Z}^{\prime} l_x + \bar{\nu}_x \slashed{Z}^{\prime}  \nu_x -  \bar{l}_y \slashed{Z}^{\prime} l_y - \bar{\nu}_y \slashed{Z}^{\prime}  \nu_y \right)  + \frac{1}{2} (m_{Z^{\prime}})^2 Z^{\prime \, \mu} Z^{\prime}_\mu \, ,
\end{eqnarray}
where $x,y =e, \mu, \tau$ and $x \neq y$. $g^{\prime}$ and $m_{Z^{\prime}}$ are the effective coupling of $Z^{\prime}$ to leptons and the mass of $Z^{\prime}$ respectively. 

We note that, in principle, there is the possibility of kinetic mixing through a $\epsilon_{xy} B_{\mu \nu} F^{\prime \mu \nu}$ term in the Lagrangian. However, this term can be forbidden at tree level through discrete symmetries\footnote{For the phenomenological implications of non-zero tree-level kinetic mixing see e.g.~\cite{Hapitas:2021ilr}.}, and then only be generated through charged-lepton loops, which for our case of interest $m_Z' > m_{l_x}, m_{l_y}$ is further suppressed by $ (m_{l_x}^2 - m_{l_y}^2) / m_Z'^2$. For details see Appendix~\ref{app.kinmix}. 

Here we remain agnostic about the mass generation mechanism for the $Z'$, and we ignore as well additional degrees of freedom present in complete models (e.g. extra scalars giving mass to the $Z'$, or new fields charged under the $U(1)$ which could be dark matter candidates, play a role in the genesis of neutrino masses, give rise to the matter-antimatter asymmetry of the universe, etc). Hence the above Lagrangian should be considered as a simplified model, having $g'$ and $m_Z^{\prime} $ as free parameters for each possible choice of $x,y$. While this gives in principle three different models, $L_e - L_{\mu}$, $L_e - L_{\tau}$ and $L_{\mu} - L_{\tau}$, we will concentrate only on the former two, for reasons that we will explain below.

\begin{figure}
    \centering
    \includegraphics[width=0.98\linewidth]{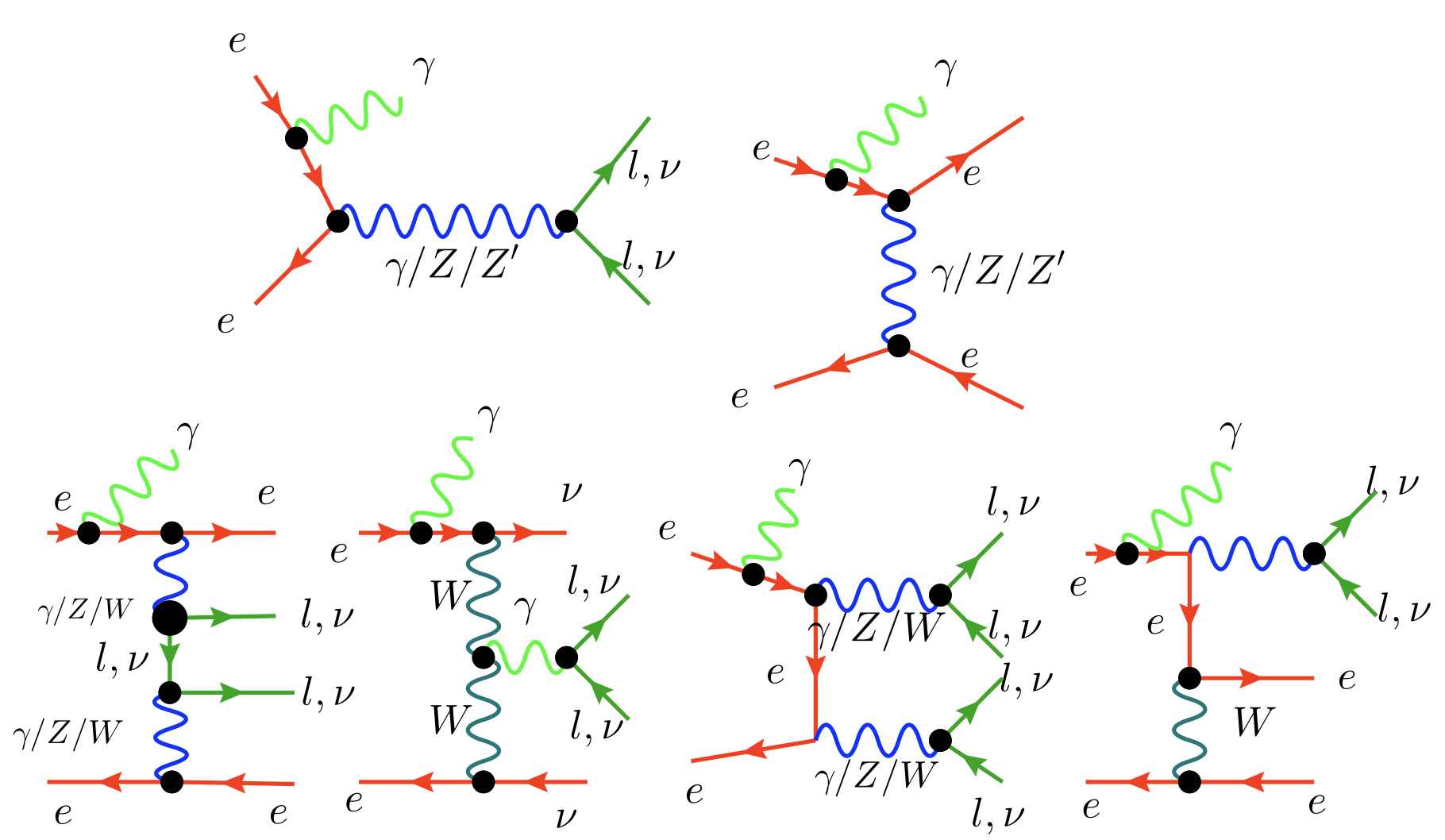}
    \caption{Representative Feynman diagrams for the $2 \to 3,4$ processes considered in this work. The upper row shows the $s-$ and $t-$ channel signals (and background through SM vector bosons). Note that the t-channel process is only contributing to $e^+ e^- \gamma$ and $\nu \nu \gamma$ final states. The lower row shows the $e^+ e^- \to \gamma 4 l$  background processes, where from the four leptons we can have either zero, two or four being charged (or neutral). }
    \label{fig:fd}
\end{figure}

First and foremost, since our objective is to study these models at $e^+ e^-$ colliders, a coupling to electrons is needed to have processes with the lowest particle multiplicity in the final state. The $e^+ e^- \to l^+ l^-$ process is the one with the lowest multiplicity, but as $Z'$ enters non-resonantly this has a lower cross section than the $e^+ e^- \to l^+ l^- \gamma$, where by making use of the radiative return the $Z'$ is produced on-shell ( see figure~\ref{fig:fd}). For the $L_{\mu} - L_{\tau}$ the lowest multiplicity process corresponds to the $2 \to 4$ process with a $Z'$ stemming from final state radiation of a $\mu$ or $\tau$ leg, which also involves an off-shell $Z'$, and hence its rate is suppressed with respect to the other two. 

Furthermore, we will restrict the mass range to $m_Z' \gtrsim 10$ GeV.\footnote{For recent work on the lighter mass range see~\cite{DeRomeri:2024dbv}.}  While we could consider accessing masses in the $[5-10]$ GeV range, there are strong constraints from Babar~\cite{Lees:2014xha} yielding $\gp \lesssim 10^{-4}$.\footnote{We have computed these constraints with the help of DarkCast~\cite{Ilten:2018crw}. In the process, we found a minor bug in the calculation of branching ratios of the leptophilic $Z'$ models, which has been reported to the DarkCast authors.} We can anticipate here that the expected sensitivity of the FCC-ee will not be able to reach those values, lying in the ballpark of $10^{-3}$. Moreover, lower masses entail additional complications in terms of modelling the reconstruction of a resonance in regions of charged lepton transverse momenta close to the reconstruction thresholds. 

Within this mass range, all possible $Z'$ decays into charged leptons and neutrinos are allowed, hence its width (ignoring neutrino masses) reads
\begin{equation}
\Gamma_{Z'} = \frac{{\gp}^2}{24 \pi} \mzp \sum_{i=x,y} \Biggl[ 2 (1+ 2 r_{i} ) \sqrt{1- 4 r_{i}} + 1 \Biggr] \approx \frac{{\gp}^2}{4 \pi} \mzp \, ,
\end{equation}
where $r_{x} = m_{l_x}^2 / \mzp^2$ and in the last term we have also ignored the charged lepton masses. In this limit, $Z'$ decays with equal branching ratios (1/3) into $l_x l_x$, $l_y l_y$, and neutrinos. From this expression, and  given that ${\gp}^2 / 4 \pi \lesssim 10^{-4}$ (as discussed in section~\ref{ss.constraints})  we can draw two important conclusions. First and foremost, that $Z'$ decays promptly.\footnote{The couplings required to have a macroscopic lifetime would give a negligible production rate at the planned FCC-ee runs.} Second, that the $m_{Z'}$ needs to differ less than 1\% from $\sqrt{s}$ for the resonant effects to start being relevant in the $2 \to 2$ process. As we have a flat prior on $\mzp$, we do not have any compelling reason why to study in detail these ``resonant'' mass ranges. All in all, this validates our strategy of focusing on $e^+ e^- \to l^+ l^- \gamma$ and $e^+ e^- \to \nu \nu \gamma$ processes at the FCC-ee.

\subsection{Existing constraints on the parameter space}
\label{ss.constraints}
\begin{figure}
    \centering
    \includegraphics[width=0.45\linewidth]{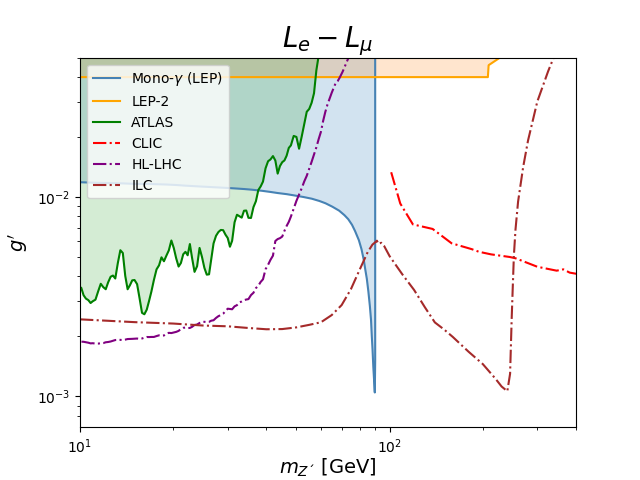}
     \includegraphics[width=0.45\linewidth]{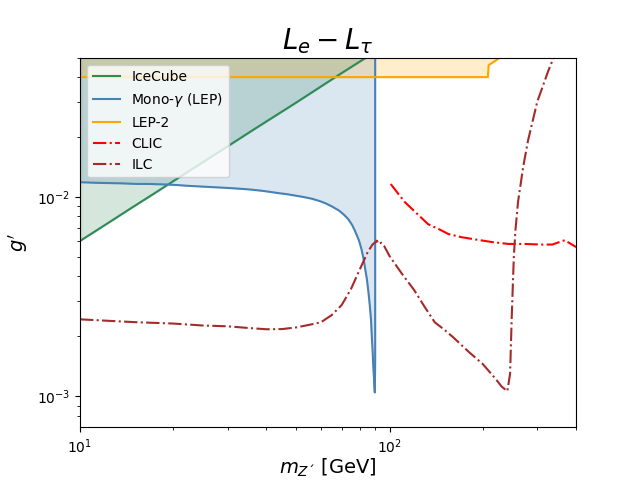}

    \caption{Constraints from current data (shaded regions, see main text for details) and the projections (dashed and dotted lines) for $e^+ e^-$ colliders operating at $\sqrt{s}=$ 3 TeV from~\cite{Dasgupta:2023zrh} and $\sqrt{s}=$ 250 GeV from~\cite{Kalinowski:2021tyr}, and the high luminosity LHC (HL-LHC) ~\cite{ATLAS:2024uvu} .}
    \label{fig:constraints}
\end{figure}

For the Lagrangian in Eq. \ref{e.lagzpl} and for the $m_{Z}^{'} \gtrsim 10$ GeV range of interest, relevant constraints come from LEP data, as well as also from LHC~\cite{CMS:2018yxg,ATLAS:2023vxg,ATLAS:2024uvu}, and from neutrino experiments. We present them here in Figure~\ref{fig:constraints}. These constraints are discussed in detail in references~\cite{Ilten:2018crw,Bauer:2018onh,Dasgupta:2023zrh} and we refer the interested reader to those publications.

For light masses, we find that the most stringent bounds are set by either 
i) LHC searches for a $Z'$ produced in a rare SM vector boson decay, $p p \to Z \to Z' \mu^+ \mu^- \to \mu^+ \mu^- \mu^+ \mu^-$ by ATLAS and CMS~\cite{CMS:2018yxg,ATLAS:2023vxg} (and more recently a search using the charged current with three muons and a neutrino in the final state)~\cite{ATLAS:2024uvu},
\footnote{These searches essentially probe the $Z'$ coupling to muons, and hence the obtained bounds are applicable to both $L_e - L_{\mu}$ and $L_{\mu} - L_{\tau}$ models, but not to $L_e - L_{\tau}$.} 
or ii) the coupling to electron neutrinos affecting the matter propagation, derived using IceCube preliminary results~\cite{IceCube:2022pbe}. We also perform a naive extrapolation of the LHC search by scaling the \emph{expected} limit assuming the HL-LHC will be statistically limited, hence assuming that the limit on the coupling scales with ${\cal L}^{1/4}$, which is shown as a dot-dashed purple line in figure~\ref{fig:constraints}.

For heavier masses, the monophoton searches at LEP become relevant, where the limits are obtained from the reintepretation~\cite{Fox:2011fx} performed of the DELPHI monophoton data~\cite{DELPHI:2003dlq,DELPHI:2008uka}, a bound that is effective up to a $Z'$ mass of about 100 GeV. For higher masses, relevant constraints arise from the study of the $e^+ e^- \to e^+ e^-$ process. Above the LEP-II center-of-mass energy of 209 GeV, effective-field theory is applicable and the bound reads $g' \geq 2.2 \times 10^{-4} \mzp$, for lighter masses a conservative bound of $g' \gtrsim 0.04$ must be used instead~\cite{Buckley:2011vc}. In what follows, we will consider the most stringent constraint for each $m_{Z}^{'}$ and only display the envelope of those limits on our plots as ``Excluded''.  It is worth noting that previous works have examined in detail the case of $\sqrt{s}=3$ TeV $e^+ e^-$ and $\mu^+ \mu^-$ colliders. Moreover, a phenomenological study on the ILC reach ($\sqrt{s}=250$ GeV) with monophotons on vector mediator of dark matter at the ILC~\cite{Kalinowski:2020lhp,Kalinowski:2021tyr}  has presented updated limits for the case of dark matter mass of 1 GeV~\cite{limitsILC}, albeit with widely different $Z'$ couplings to electrons and dark matter. We performed a reinterpretation of the expected limits of the search into our model parameters.\footnote{We simply assumed that the sensitivity scales with $(g^{\prime})^2 \times {\rm BR} (Z^{\prime} \to \nu \nu )$, and rescaled the results from a 100 \% invisible branching fraction to our values, which go asymptotically to $1/3$ when lepton masses can be neglected. We thank Filip Zarnecki for reminding us about this study, for sharing their unpublished limits in electronic format with us, and for useful discussions on how to accurately reinterpret them.} We incorporate these future collider projections as dot-dashed red and brown lines for CLIC and ILC respectively in figure~\ref{fig:constraints}.

From the figure it is clear that there is a fertile territory in parameter space that the FCC-ee, in particular the low mass region, $\mzp \sim 10-100$ GeV, where CLIC and the HL-LHC start to lose sensitivity.

Finally, as the proton does have a small lepton parton distribution function~\cite{Manohar:2017eqh,Buonocore:2020nai}, it is important to verify how LHC bounds would apply by obtaining one electron and one positron (or muons, or taus) from the proton pair. Implementing the \textsc{LUXPDF} formalism from~\cite{Buonocore:2020nai} in \textsc{MG5}\_aMC@NLO~3.5.1~\cite{Alwall:2014hca}, we find that the LHC cross section for our search channel $e^+ e^- \to Z^{\prime}$ is of the order $10^{-(4\div6)}$ pb for couplings $g \sim 0.1$, which agrees with the estimations made in~\cite{Buonocore:2020nai}. This yields only a handful of signal events for relatively low masses at the HL-LHC, and hence we will ignore it in what follows.

\section{Future Circular Collider in the $e^+ e^-$ mode} 

The Future Circular Collider (FCC) at CERN is a proposed collider that would operate in two stages: FCC-ee, a high-luminosity electron-positron collider; followed by FCC-hh, a hadron collider of ultra-high energy. It will be housed in a 90~km long tunnel at CERN and will accommodate 4 interaction points. The first stage, FCC-ee \cite{Agapov:2022bhm, Bernardi:2022hny}, will provide ultimate precision measurements of the Higgs boson, top quark, and electroweak interactions. FCC-ee will operate at various collision energies, ranging from the Z pole to the top quark pair production threshold, spanning a 16-year research program. This is a crucial step towards exploring physics beyond the Standard Model and will pave the way for the subsequent hadron collider which aims to reach collision energies of 100 TeV and above.
The considered beam energies and total integrated luminosities, corresponding to the proposed runs of the FCC-ee, are presented in Table~\ref{t.runs}. 

\begin{table}[!htp]
\centering
\begin{tabular}{ccc}
\hline
\multicolumn{1}{l}{Run Name} & \multicolumn{1}{l}{$E_{\rm beam}$ {[}GeV{]}} & \multicolumn{1}{l}{$\int {\cal L}$ {[}ab$^{-1}${]}} \\ \hline
Z          & 45.6    & 205                                                  \\
WW         & 80      & 10                                                   \\
ZH         & 120     & 7.2                                                  \\
tt       & 182.5          & 2.68                                                
\end{tabular}
\caption{Beam energies and total integrated luminosities considered in our study. Taken from Table 129, FCC-mid-term report~\cite{FCCMidTerm}. }
\label{t.runs}
\end{table}

\section{Collider analysis}
\label{s.coll}
We perform our collider analysis using MadGraph5\_aMC@NLO v 3.5.1 to generate events at the parton level. Subsequently, the events are showered using PYTHIA8 \cite{Bierlich:2022pfr}. The detector effects are taken into account by running DELPHES \cite{deFavereau:2013fsa} with the IDEA card. We employ MadAnalysis5~\cite{Conte:2012fm,Conte:2014zja,Dumont:2014tja,Conte:2018vmg,Araz:2019otb,Araz:2020lnp} to process the reconstructed events.

For the search channel under consideration, the signal process $e^+ e^- \to \gamma Z^{\prime}$, with the $Z^{\prime}$ further decaying to a lepton pair ($e^+ e^-$, $\mu^+ \mu^-$, $\tau^+ \tau^-$ or a $\nu$ pair of desired flavour) is generated using the UFO model file \textsc{DMSimp}~\cite{Backovic:2015soa,Albert:2017onk}. The \textsc{DMSimp} model is a simple extension of the SM with a few new fields. The vector DM particle \texttt{y1}, is our $Z^{\prime}$ vector boson. We kinematically forbid the decay of the $Z^{\prime}$ to the dark sector particles of the model by setting the mass of the DM particles an order of magnitude higher than the highest possible $Z^{\prime}$ mass in our region of interest. The possible backgrounds for this search channel are the SM counterparts of the signal processes, viz. $e^+ e^- \to \gamma l l$ (where $l$ stands for all possible charged and neutral leptons) and the vector-boson fusion induced four-lepton final states: $e^+ e^- \to l_x l_x l_y l_y \gamma$, where $l_x, l_y$ are all possible lepton flavours. The sensitivity of our detector depends largely on its specifications and on the object definitions. 

 For the signal processes, we generate 10000 events corresponding to a $g^{\prime}$ of 0.1, for mass increments of 5 GeV up to the maximum kinematically allowed mass. This gives us the cross-sections $\sigma_{0.1}$ for each mass point. The coupling scales as ${g^{\prime}}^{2}$, hence the cross-section $\sigma$ for any given coupling $g^{\prime}$ is given by: 
\begin{equation}
\label{eq:3}
\sigma = \sigma_{0.1} \times \left(\frac{g^{\prime}}{0.1}\right)^{2} \, .
\end{equation}

We impose, guided by the Delphes IDEA card, the following preselection cuts
\begin{itemize}
\item{$l= e, \mu$: $p_T > 3$ GeV, $|\eta| \leq 2.56$, $\Delta R(l,X)$ > 0.5,  }
\item{$\gamma$: $E > 2 $ GeV, $p_T > 1$ GeV,  $|\eta| < 3.0$,  $\Delta R(\gamma,X) >$ 0.5, }
\item{$\tau$: $p_T > 1$ GeV,  $|\eta| \lesssim 3.0$, $\Delta R(\tau,X)$ > 0.5 .}
\end{itemize}
 We have employed looser cuts at the parton level, and explored the consequences of relaxing some the $p_T(l), E(\gamma)$ cuts, finding that they do not affect the signal samples in the mass range of interest.  Table~\ref{t.events} lists the expected number of background events for each of the FCC-ee runs following the parton-level analysis.

\begin{table}[!htp]
\centering
\begin{tabular}{ccccc}
\hline
        \multicolumn{1}{|l|}{Process }                & \multicolumn{4}{c|}{N$_{ev}$}                                                                                                                    \\ \hline
\multicolumn{1}{|l|}{$e^+ e^- \to \gamma +...$} & \multicolumn{1}{l|}{Z run} & \multicolumn{1}{l|}{WW run} & \multicolumn{1}{l|}{ZH run} & \multicolumn{1}{l|}{$t\bar{t}$ run}                    \\ \hline

$\mu \mu$          & $2.3 \times 10^{10}$      & $2.1 \times 10^{7}$   & $5.5 \times 10^{6}$    & $8.44 \times 10^{5}$                \\
$e e$              & $8.63 \times 10^{10}$     & $1.26 \times 10^{9}$  & $4.5 \times 10^{8}$    & $7.9 \times 10^{7}$   \\
$\tau \tau$        & $2.3 \times 10^{10}$      & $2.1 \times 10^{7}$   & $5.7 \times 10^{6}$    & $8.82 \times 10^{5}$                 \\
$\nu \nu$          & $2.2 \times 10^{9}$      & $5.9 \times 10^{7}$   & $3.3 \times 10^{7}$    & $1.35 \times 10^{7}$  \\
$\mu \mu \mu \mu$  & $1.2 \times 10^{5}$      & $1.4 \times 10^{4}$                  & $6.3 \times 10^{3}$                  & $1.4 \times 10^{3}$    \\
$\mu \mu e e$      & $8 \times 10^{7}$        & $5.03 \times 10^{6}$  & $4.16 \times 10^{6}$   & $1.73 \times 10^{6}$ \\
$\mu \mu \tau \tau$& $1.43 \times 10^{9} $   & $9.9 \times 10^{6}$ & $1.7 \times 10^{8}$     & $2.3 \times 10^{6}$  \\
$\mu \mu \nu \nu$  & $8 \times 10^{3}$                      & $1.8 \times 10^{4}$                  & $1.56 \times 10^{4}$     & $6.7 \times 10^{3}$                          \\
$e e e e$          & $7.6 \times 10^{7}$      & $4.86 \times 10^{6}$  & $4.04 \times 10^{6}$   & $1.78 \times 10^{6}$                   \\
$e e \tau \tau$    & $3 \times 10^{7}$         & $1.1 \times 10^{6}$      & $8.9 \times 10^{5}$      & $3.82 \times 10^{5}$              \\
$ e e \nu \nu$     & $1.28 \times 10^{4}$                    & $2 \times 10^{4}$                     & $2.5 \times 10^{4}$       & $1.16 \times 10^{4}$               \\
$\tau \tau \tau \tau$         & $5 \times 10^{5}$       & $6.3 \times 10^{3}$      & $4.5 \times 10^{3}$      & $1 \times 10^{3}$                                          \\
$\tau \tau \nu \nu$         & $4 \times 10^{3}$         & $2.3 \times 10^{4}$        & $1.6 \times 10^{5}$      & $4 \times 10^{4}$                                         \\
$\nu \nu \nu \nu$         & $0.5$        & $1.12 \times 10^{4}$     & $8.1 \times 10^{3}$      & $4.6 \times 10^{3}$                                       
\end{tabular}
\caption{Expected number of events for our background processes, for the four runs considered in this study.   }
\label{t.events}
\end{table}

We will perform four separate studies, for the $l=e,\mu,\tau,\nu$ leptons. For the charged leptons, we request to have at least a pair of charged leptons of a given flavour, and veto all other charged lepton flavours. For example, for the $e^+ e^- \gamma$ final state we request $N_e \geq 2, N_{\mu, \tau} = 0$. For the invisible final state, we veto events with any number of reconstructed charged leptons.

\begin{figure}
\centering
\includegraphics[width=1\textwidth]{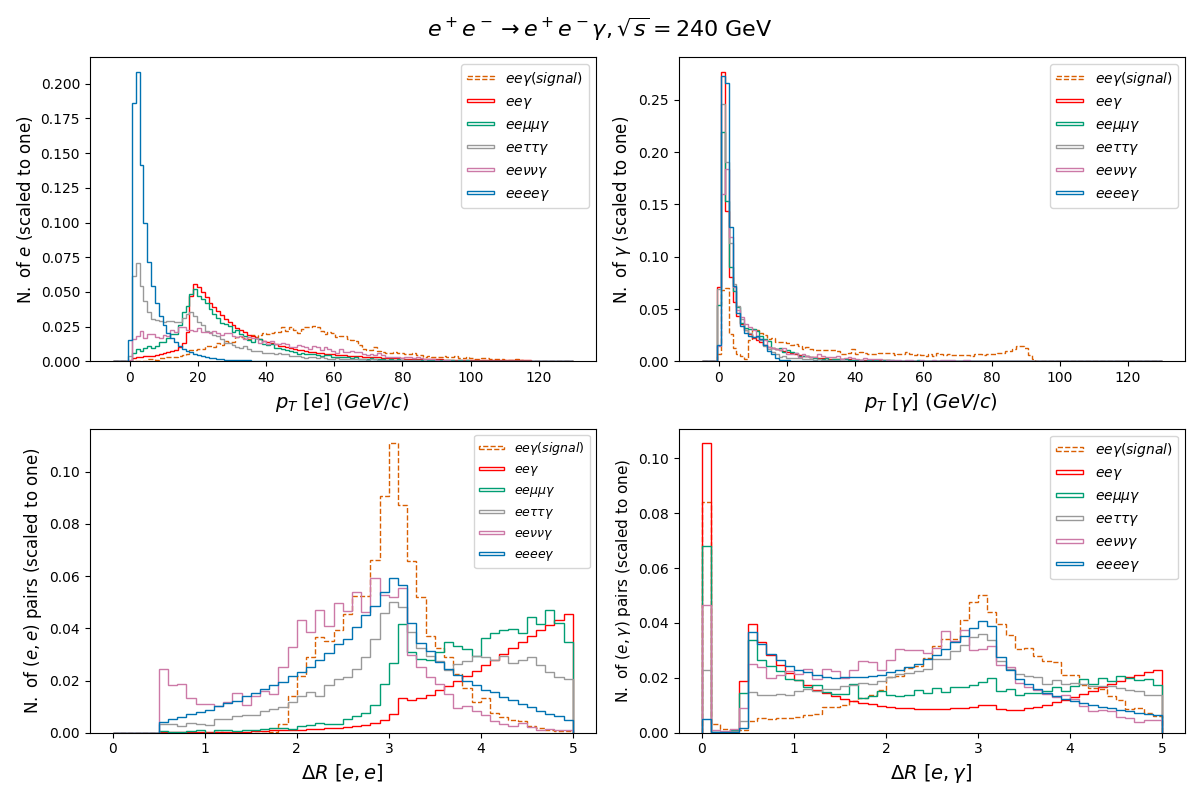}
\caption{Normalized kinematic distributions of signal and background events for the $ee\gamma$ search channel at the ZH run with $m_{Z'}$=120 GeV, $\gp$= 0.1. The distributions shown are (from left to right, top to bottom): transverse momentum $p_{T}$ of electrons, $p_{T}$ of photons, $\Delta R$ of electron pairs and $\Delta R $ of a photon and an electron.
}
\label{fig:distribution_4}
\end{figure}

\begin{figure}
\centering
\includegraphics[width=1\textwidth]{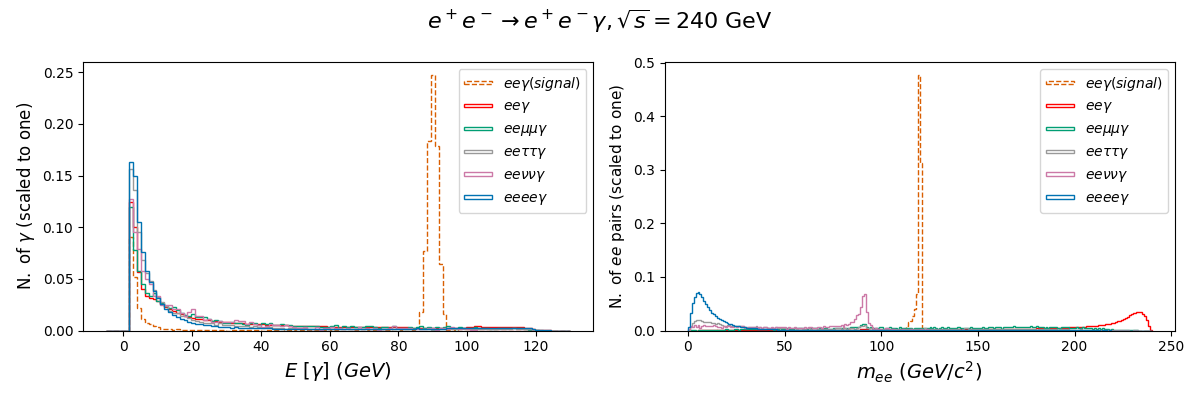}
\caption{Normalized kinematic distributions of signal and background events for the $ee\gamma$ search channel at the ZH run with $m_{Z'}$=120 GeV, $\gp$= 0.1. The plot on the left shows the photon energy distribution, and the plot on the right shows the invariant mass distribution of the electron pair.}
\label{fig:distribution_2}
\end{figure}

To illustrate the event kinematics, we show in Figs.~\ref{fig:distribution_4} and ~\ref{fig:distribution_2} the normalized distributions of the events that survive the above selection procedure, for the $ee\gamma$ search. The plots are shown for $m_{Z^{\prime}}=120$ GeV and $\gp =$ 0.1, at the ZH run ($\sqrt{s}$= 240 GeV). From the considered variables, it is clear that the maximum discrimination between signal and background comes from the invariant mass $m_{ee}$, and $E_{\gamma}$ distributions. In the case of $m_{ee}$, the four-lepton backgrounds $ee\mu\mu\gamma$ and $ee\tau\tau\gamma$ are highly suppressed with a flat distribution, with around $10^{-3}$ events per bin. 

In the $E_{\gamma}$ distribution, we see a peak in the photon energy that has a value given by 
\begin{equation}
E_{peak} =  \frac{m_{Z^{\prime}}^{2}}{2\sqrt{s}} - \frac{\sqrt{s}}{2} \, .
\end{equation}
However, given the broadness of the $E_{\gamma}$ peak compared to the $m_{ee}$ one, we can anticipate that the sensitivity obtained from the invisible channel would be smaller than that derived from the charged lepton pair.\footnote{The IDEA card we employed assumes a fiber electromagnetic calorimeter, with a stochastic term in the resolution of 11 \%, while the alternative crystal calorimeter assumes an stochastic term of 3 \%. Hence our analysis in the invisible channel is conservative. We thank Giacomo Polesello for the clarification.} 

Hence our study proceeds by examining the invariant mass distribution of the di-lepton pairs and the recoil mass. We define the invariant mass window $\Delta_{ll}$ and the energy window $\Delta_{\gamma}$ as follows
\begin{equation}
|m_{ee} - \mzp| < \Delta_{ll} , \, |E_{\gamma} - E_{peak}| < \Delta_{\gamma} \, .
\end{equation}
The sensitivity of our search is characterized by the significance $\mathcal{Z}$: 
\begin{equation}
\label{eq:1}
\mathcal{Z}=\frac{N_s}{\sqrt{N_s+N_b + \lambda^2 N_b^2}} ,
\end{equation}
where $N_s$ and $N_b$ are the number of signal events and background events respectively, in the 
signal region under consideration, and $\lambda$ is the systematic error. In this work, we present sensitivities for a value of significance $\mathcal{Z}=2$, which corresponds to a $95\%$ CL. Unless otherwise noted, we will ignore systematics ($\lambda = 0$).

\begin{figure}
\centering
\includegraphics[width=0.49\textwidth]{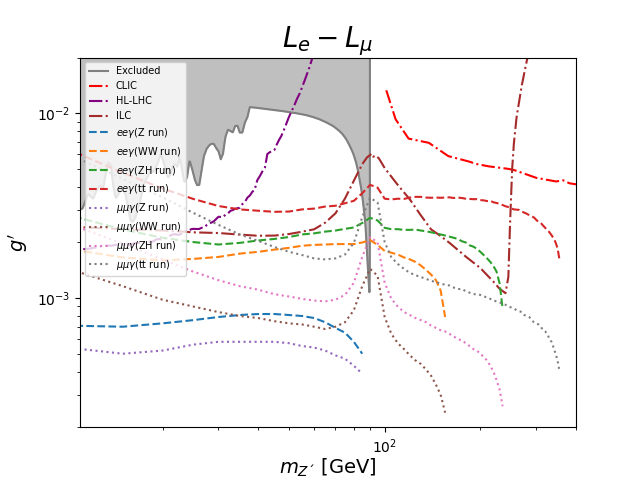}
\includegraphics[width=0.49\textwidth]{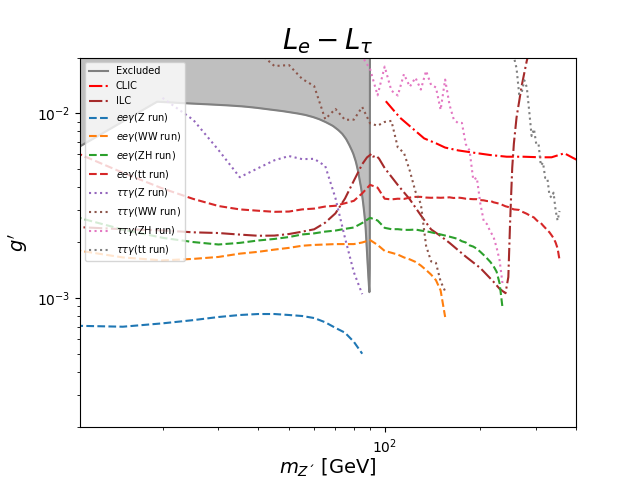}
\caption{Projected sensitivities of the four FCC-ee runs to the $L_e - L_\mu$ (left) and $L_e - L_\tau$ (right) models, for the charged dilepton pair search channel. The dashed lines correspond to the sensitivities of the four FCC-ee runs for the $ee\gamma$ search channel. The dotted lines show the reach of the $\mu \mu \gamma$ (left) and $\tau\tau\gamma$ (right) searches. The existing constraints are displayed as a grey filled area, while dot-dashed lines show the projected reach of the HL-LHC (purple), CLIC (red) and ILC (brown). See main text for details.  }
\label{fig:sensitivitycharged}
\end{figure}

Our sensitivity projections for $\Delta_{ll} =$ 5 GeV are shown in 
Fig.~\ref{fig:sensitivitycharged}, for the $L_e - L_{\mu}$ (left panel) and the $L_e - L_{\tau}$ (right panel) models.
From the figure we see that among the three possible charged leptons, muons perform better than electrons, that in turn perform better than taus. As luminosity is a key factor, the $Z$ run is the most sensitive one, reaching values of about $5 \times 10^{-4}$ for the $\gp$ coupling in almost all the mass range. Near the $Z$ peak, it is clear that the search loses sensitivity due to the background from the Z-boson production. It is also worth noting that the muon and electron final states outperform the expected CLIC reach from~\cite{Dasgupta:2023zrh}. We see that, except for the Z-boson mass region, this single study increases the existing limits on $\gp$ by one to two orders of magnitude.

\begin{figure}
\centering
\includegraphics[width=0.49\textwidth]{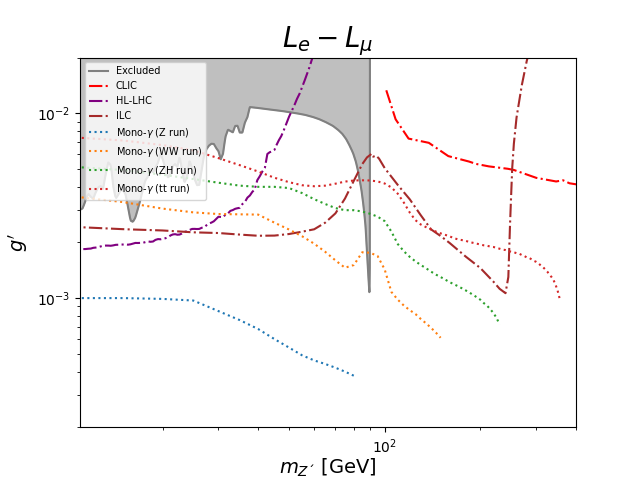}
\includegraphics[width=0.49\textwidth]{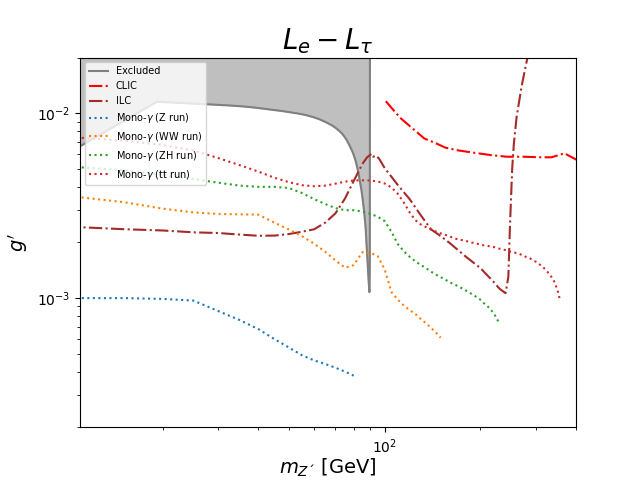}
\caption{Projected sensitivities of the four FCC-ee runs to the $L_e - L_\mu$ (left) and $L_e - L_\tau$ (right) models, for the mono-photon search channel (dotted lines). The existing constraints are shown in a gray shadowed area, while dot-dashed lines show the projected reach of the HL-LHC (purple) and CLIC (red). See main text for details. }
\label{fig:sensitivitymonophoton}
\end{figure}

In figure~\ref{fig:sensitivitymonophoton} we examine instead the sensitivity of the ``mono-photon'' final state, using a reference value of $\Delta_{\gamma} =  5$ GeV.
This search channel, on its own, is weaker than the $\mu$ and $e$ channels.However, it is still an important improvement over the existing limits. It is also worth noting that in our setup the branching fractions into visible and invisible are correlated, but in more general constructions with additional freedom this channel could play an important role (for instance if the $Z'$ would decay to dark sector particles).

\begin{figure}
\centering
\includegraphics[width=0.95\textwidth]{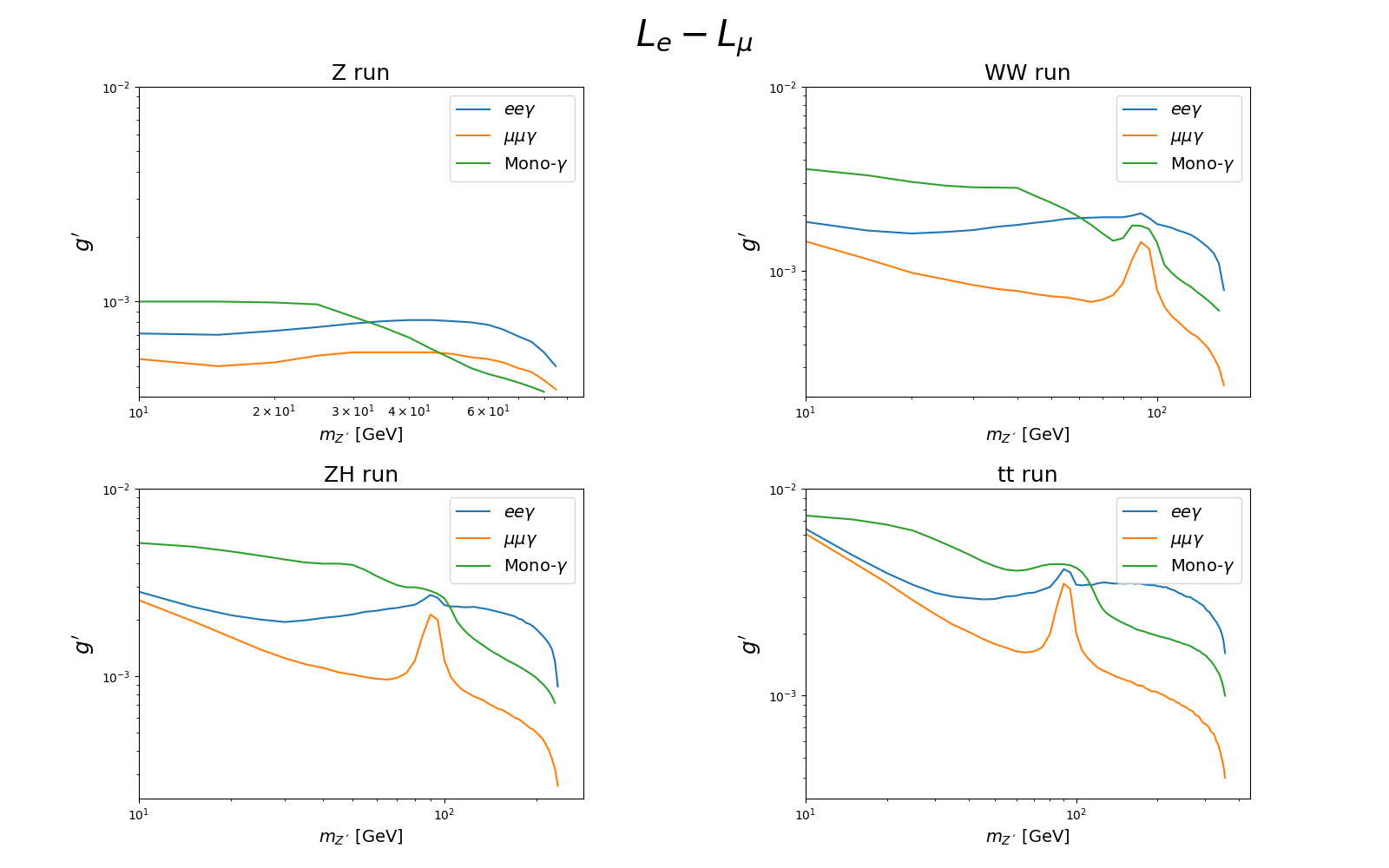}
\caption{Projected sensitivities of the four FCC-ee runs to the  $L_e - L_{\mu}$ model for the $ee \gamma$, $\mu \mu \gamma$ and mono-photon channels.}
\label{fig:sensitivityall_emu}
\end{figure}

\begin{figure}
\centering
\includegraphics[width=0.95\textwidth]{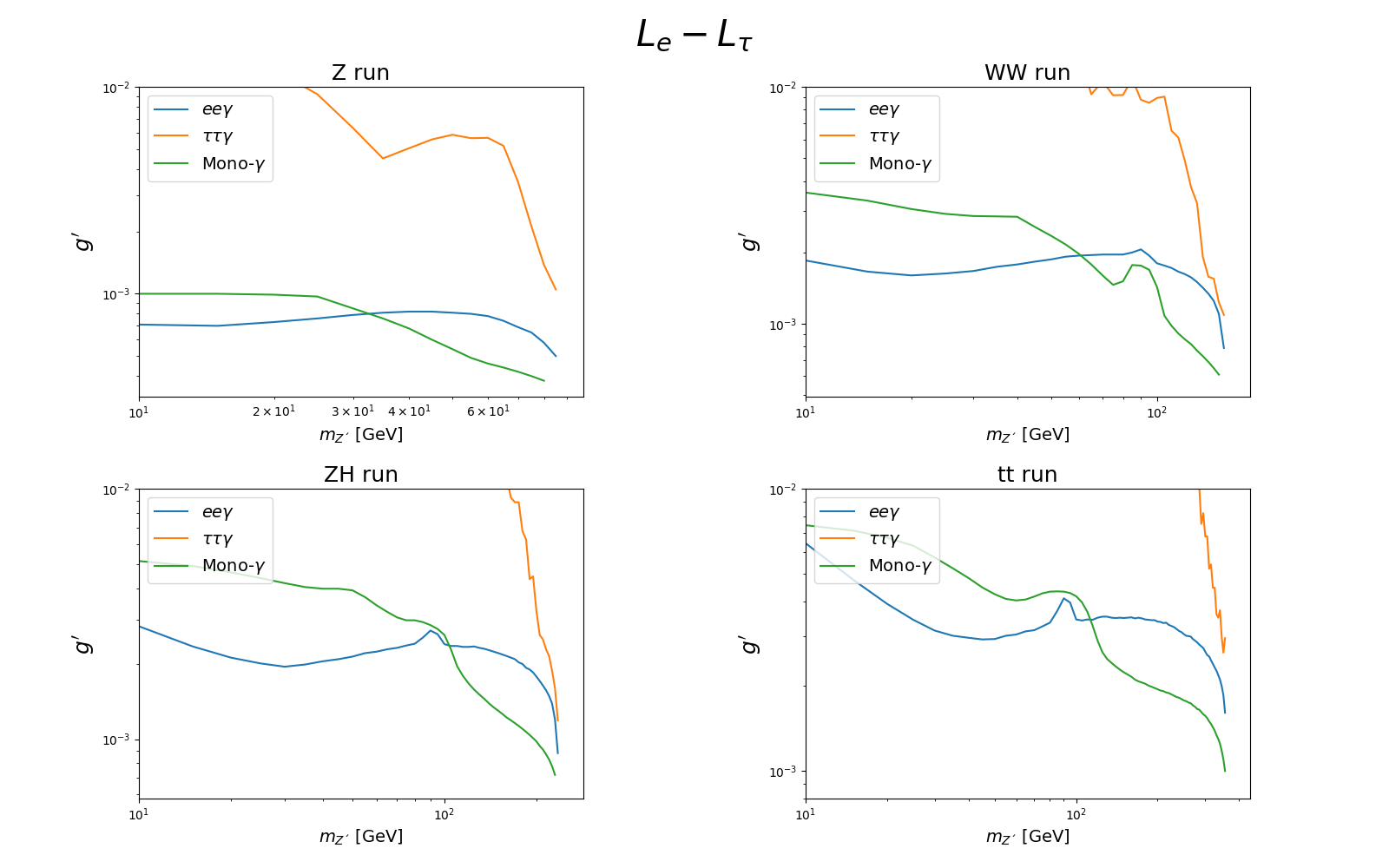}
\caption{Projected sensitivities of the four FCC-ee runs to the $L_e - L_{\tau}$ model for the $ee \gamma$, $\tau \tau \gamma$ and mono-photon channels.}
\label{fig:sensitivityall_etau}
\end{figure}

For better visualization, we also display the individual sensitivities from each search channel, for each planned FCC-ee run, for the $L_e - L_{\mu}$ and $L_e - L_{\tau}$ in figures~\ref{fig:sensitivityall_emu} and ~\ref{fig:sensitivityall_etau}, respectively.

In figure~\ref{fig:sensitivityall_emu} we see that the $\mu \mu \gamma$ run is the most sensitive one, with the difference to the $e e \gamma$ channel shrinking with higher center-of-mass energy. We also see that close to the threshold, where the photon becomes softer, the mono-$\gamma$ channel can be the leading probe.

On the other hand, in figure~\ref{fig:sensitivityall_etau} we see that in several runs the monophoton can give the leading constraints, and as anticipated the taus have worse bounds owing to their poor reconstruction.

\begin{figure}
\centering
\includegraphics[width=0.95\textwidth]{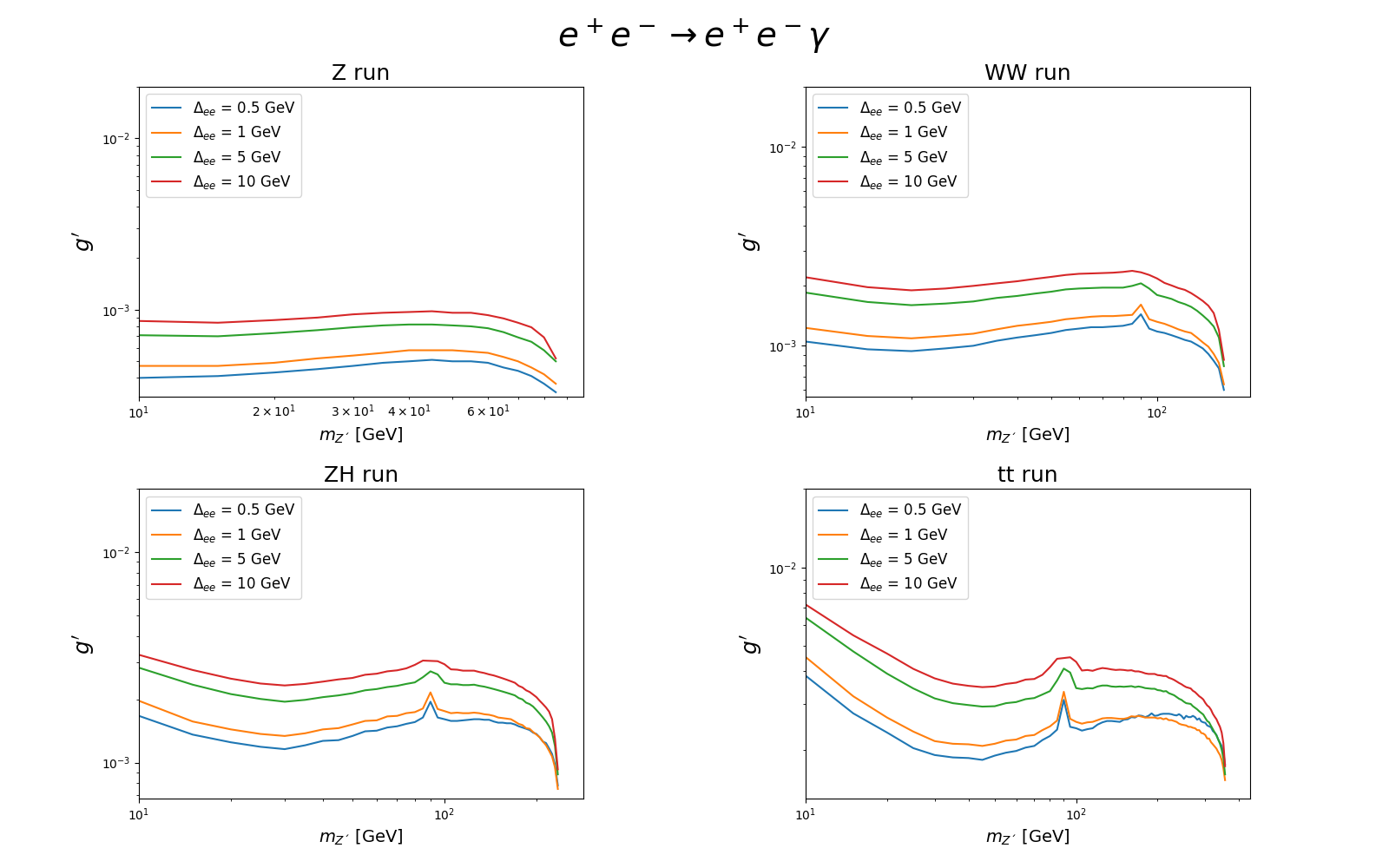}
\caption{Dependence of the FCC-ee sensitivity on $\Delta_{ll}$, shown for the $ee\gamma$ channel. The sensitivities are plotted for $\Delta_{ee}$ = 0.5 GeV, 1 GeV, 5 GeV and 10 GeV.}
\label{fig:masswindow}
\end{figure}
\begin{figure}
\centering
\includegraphics[width=0.95\textwidth]{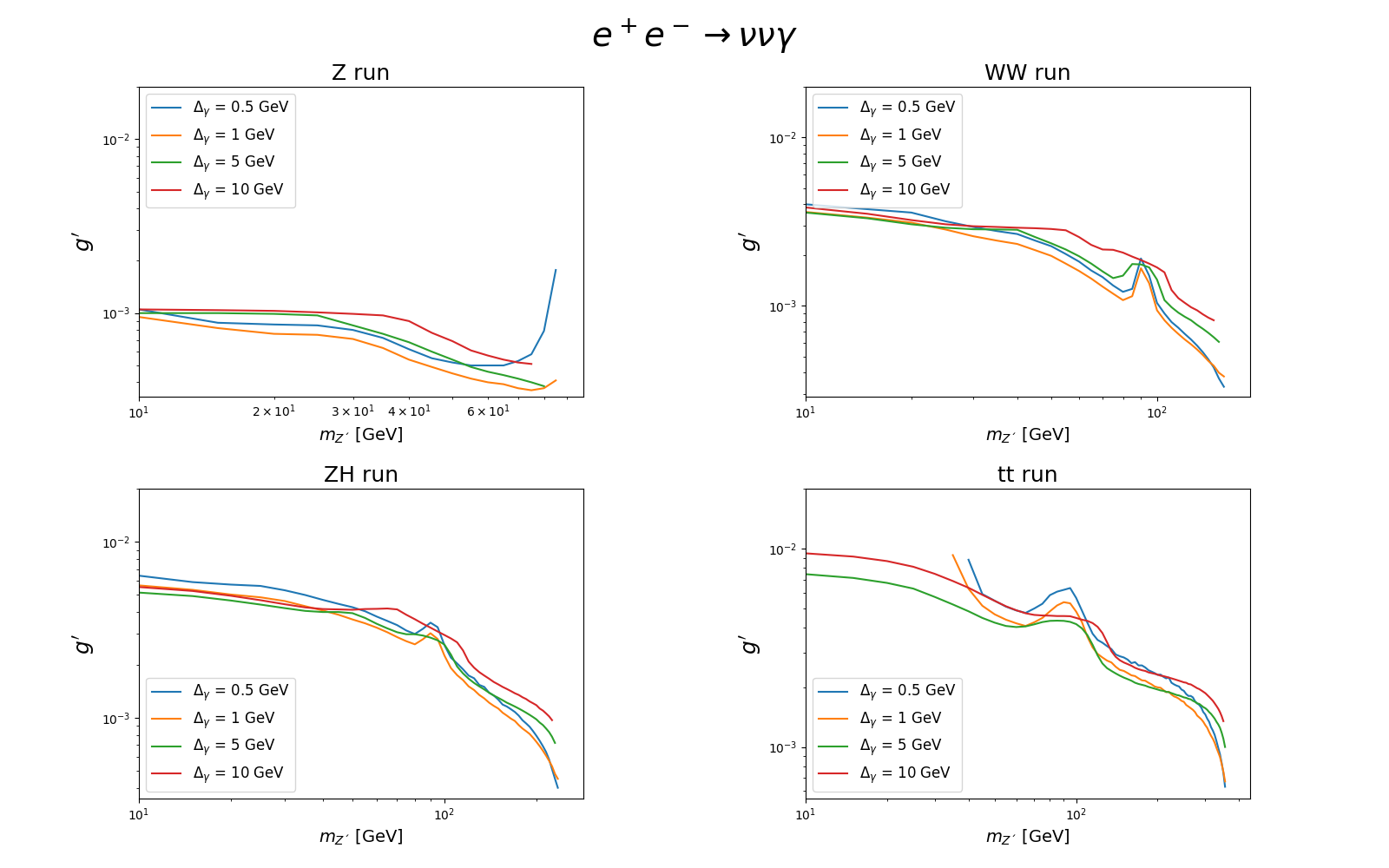}
\caption{Dependence of the FCC-ee sensitivity on $\Delta_{\gamma}$, shown for the mono-photon channel. The sensitivities are plotted for $\Delta_{\gamma}$ = 0.5 GeV, 1 GeV, 5 GeV and 10 GeV.}
\label{fig:energywindow}
\end{figure}

It is important to explore how lower value of $\Delta_{ll}$ and of $\Delta_{\gamma}$ affect the sensitivity projects. To that end, in figure~\ref{fig:masswindow} we show how the exclusion on the $ee\gamma$ final state varies for for different values of $\Delta_{ll} = 0.5, 1, 5, 10$ GeV, while in figure~\ref{fig:energywindow} we show the impact of choosing $\Delta_{\gamma} = 0.5, 1, 5, 10$ GeV.

We see that the reduction of the mass window greatly affect the coupling exclusion, accounting for a reduction of about 50 \% in the excluded $\gp$ value. For the energy window, the broader photon peak results in a milder impact of the choice of $\Delta_{\gamma}$ on the final excluded coupling. Moreover, for tighter mass windows we start cutting off the signal peak, and hence losing some sensitivity. 

\begin{figure}
\centering
\includegraphics[width=0.47\textwidth]{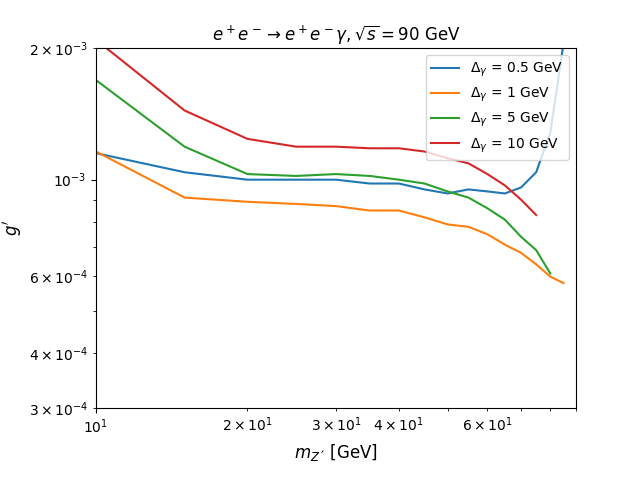}
\includegraphics[width=0.5\textwidth]{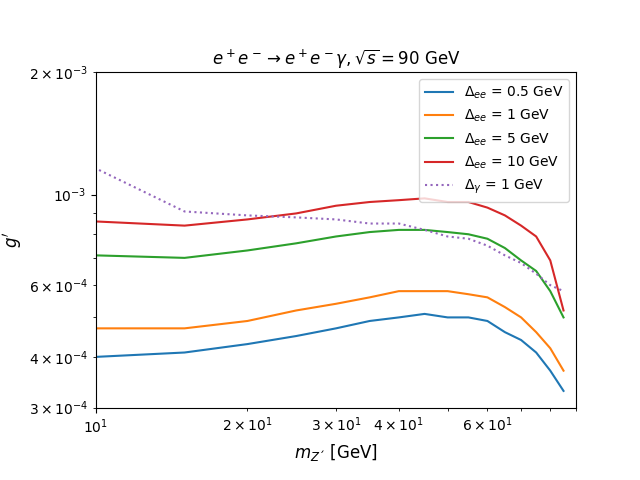}
\caption{Figure on the left shows bounds obtained for the $ee\gamma$ search channel by using the energy of the ISR photon, shown for varying energy windows. Figure on the right is a comparison of projected sensitivity of the FCC-ee to the $L_e - L_\mu$ model in the $ee\gamma$ search channel, obtained by analyzing the invariant mass window to that obtained from the energy distribution of the ISR photon. The solid lines show the variation of sensitivity with the invariant mass window. The dotted line shows the strongest constraint obtained by using the energy window of the ISR photon, corresponding to $\Delta_{\gamma}$ = 1 GeV.}
\label{fig:ll_energywindow}
\end{figure}

It is interesting to see, for the charged lepton analysis, the impact of cutting on $E_{\gamma}$ instead of on the charged lepton pair. We illustrate this for the $Z$ run in figure~\ref{fig:ll_energywindow}. In the left panel, we show the coupling exclusions for different choices of the $\Delta_{\gamma}$ variable, while fixing $\Delta_{ll}= 5$ GeV. The turnover when going from 1 GeV to 0.5 GeV is clearly visible. Using then the ``optimal'' value of $\Delta_{\gamma}$ = 1 GeV, in the right panel we overlay that sensitivity with those obtained from the charged-lepton pair analysis. We see that for low masses the energy window performs worse than any $\Delta_{ee}$ cut, and with increasing mass it performs at best as well as the $\Delta_{ee} = 5$ GeV or $\Delta_{ee} = 10$ GeV choices. Hence in what follows we will ignore the energy window and focus only on varying the $\Delta_{ee}$ value.

\begin{figure}
\centering
\includegraphics[width=0.49\textwidth]{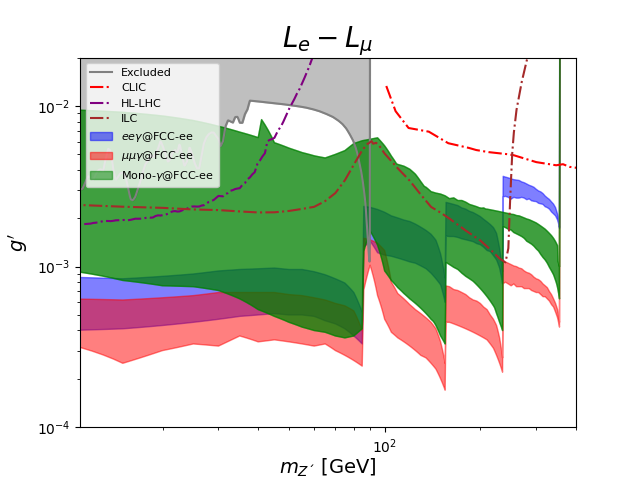}
\includegraphics[width=0.49\textwidth]{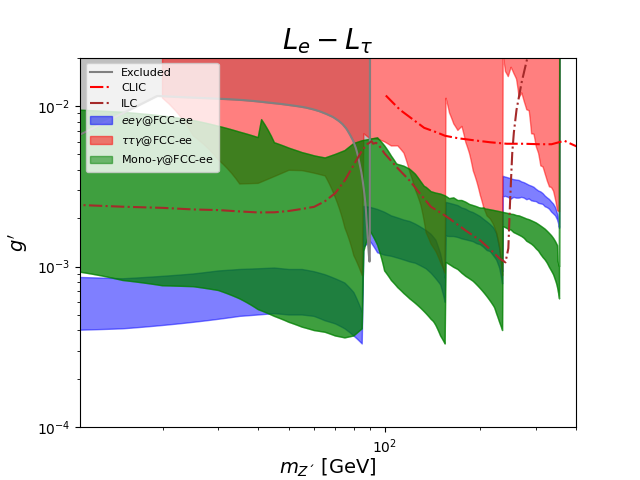}
\caption{Projected sensitivities of all four FCC-ee runs to the $L_e - L_\mu$ ($L_e - L_\tau$) model. The red, blue and green bands correspond to the $\mu\mu\gamma$ ($\tau\tau\gamma$), $ee\gamma$ and mono-$\gamma$ search channels respectively, and show the range of bounds set on this model by varying $\Delta_{ll}$ and $\Delta_{\gamma}$ for each search channel from 0.5 GeV to 10 GeV.}
\label{fig:band}
\end{figure}

Our projected sensitivities for each model, with varying $\Delta_{ll}$ (for the charged leptons) and $\Delta_{\gamma}$ (for the invisible channel) are displayed in figure~\ref{fig:band}. 
The band shows the best and worst case scenarios over the variations on $\Delta_{ll}$ and $\Delta_{\gamma}$. For low masses in the 10-20 GeV range, we see that even in the most conservative case there is a factor of two increase from the HL-LHC expectation, which increases for heavier masses up to about 80 GeV. When reaching the $Z$-pole, there is a competition with the LEP results, and after the $Z$ pole we clearly see that FCC-ee covers a large amount of ``terra incognita'', even surpassing the projected CLIC reach from~\cite{Dasgupta:2023zrh} by a factor of 2-3 in the most conservative scenario, and extending up to one-two orders of magnitude. 

\begin{figure}
\centering
\includegraphics[width=0.49\textwidth]{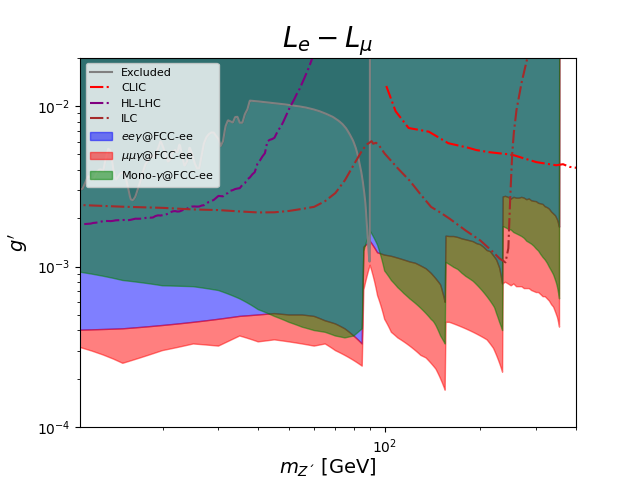}
\includegraphics[width=0.49\textwidth]{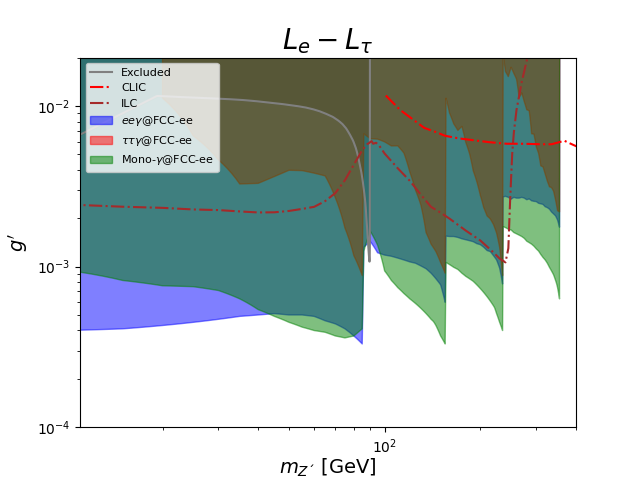}
\caption{Projected sensitivities of the FCC-ee to the $L_e - L_\mu$ ($L_e - L_\tau$) models. The red, blue and green bands correspond to the $\mu\mu\gamma$ ($\tau\tau\gamma$), $ee\gamma$ and mono-$\gamma$ search channels respectively, and show the strongest bounds set on this model by each search channel.}
\label{fig:bandoptimistic}
\end{figure}

Our most optimistic projections are then summarized in figure~\ref{fig:bandoptimistic}, filling in the parameter space that would be ultimately excluded, which is about $\gp \sim 3 (4) \times 10^{-4}$ for the $L_e - L_{\mu}$ ($L_e - L_{\tau}$) scenarios, except or the $Z-$ pole region where the limit is weaker, reaching $10^{-3} (2) \times 10^{-3}$. We note, nonetheless, that we have not attempted to combine the individual search channels (mostly because they are highly correlated), which would yield a better sensitivity. Hence a combination could compensate for the  potential loss if our choices of  $\Delta_{ll} = 0.5$ GeV and $\Delta_{\gamma}$= 1 GeV were not realistic.

\begin{figure}
\centering
\includegraphics[width=0.49\textwidth]{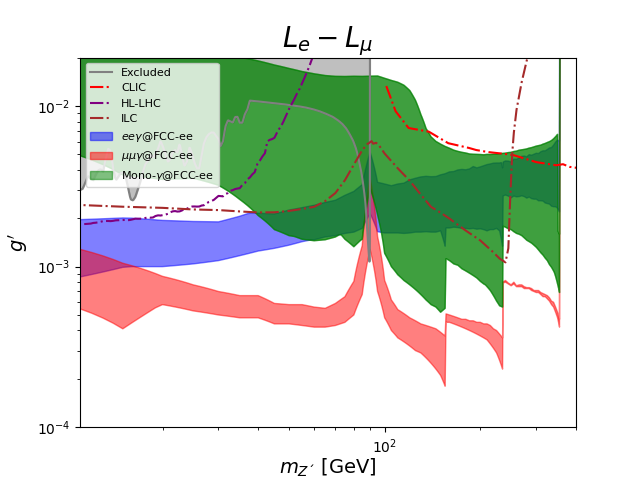}
\includegraphics[width=0.49\textwidth]{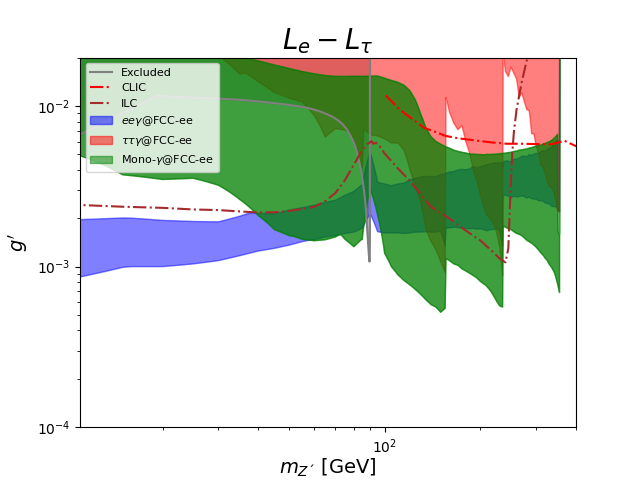}
\caption{The impact of systematics on the strongest bounds of the FCC-ee for the $L_e - L_\mu$ ($L_e - L_\tau$) models. The red, blue and green bands correspond to the $\mu\mu\gamma$ ($\tau\tau\gamma$), $ee\gamma$ and mono-$\gamma$ search channels respectively, and show the range of exclusions obtained by introducing $\lambda$ values of 0.1\% and 1\% on the constraints shown in Figure \ref{fig:bandoptimistic}.}
\label{fig:systematics}
\end{figure}

Finally, we analyze the impact of systematic uncertainties on the expected sensitivities. We show in figure~\ref{fig:systematics} the variation of the sensitivity for $\lambda \in [0.1 - 1] \%$ range, with the optimized cuts obtained from the $\lambda=0$ case.   We see that the $\tau \tau$ and $\nu \nu$ channels are highly affected by the systematic error, while for electrons the impact is mild (about a factor of 2) and for $\mu \mu$ there is no much impact from going to 0.1 to 1 \%. This provides an important target accuracy for the FCC-ee design, which given the expected clean environment of the FCC-ee (compared to the LHC) we consider our expectations for $\lambda$ to be achievable. 

\section{Conclusions}
\label{s.conclu}
Leptophilic models are an ideal target for future lepton colliders, as there is a large parameter space to be explorer. As an example of a leptophilic setup, we have considered the simple case of a new neutral vector boson ($Z'$) gauging the $L_e - L_{\mu}$ and $L_e - L_{\tau}$ group. This model has two free parameters, the coupling $g'$ and the $Z^{\prime}$ mass, hence making it an ideal benchmark to compare the capabilities of the different lepton colliders option under consideration. Indeed, as a thorough study was carried out for the CLIC and Muon Collider, here we filled in a gap and studied the FCC-ee sensitivity.

We have focused on the four planned FCC-ee runs at the $Z$ pole, and at the $WW, ZH$ and $t \bar{t}$ thresholds. We have studied in detail the final states with $l^+ l^- \gamma$, $l=e, \mu, \tau$ and $\nu \nu \gamma$, including fast detector simulation using the Delphes IDEA card with a fiber calorimeter. We have included backgrounds with a photon and up to four leptons (charged or neutral). We pursued a conservative route of exploring varying windows in the $m_{ll}$ (for the charged lepton case) and $E_{\gamma}$ (for the invisible final state). In those distributions, unless one is close to a SM resonance, the background is flat and the signal is a sharp peak on these variables. We have explored the sensitivity to different size of the mass and photon energy window. 

We have also explored the impact of systematic uncertainties, finding that they have a mild impact on the electron, tau and $\nu$ channel up to about 0.1 \%, but those channels are affected when a level of 1\% systematics is reached. In contrast, the muon channel is quite robust against systematic at the 1 \% level.

In summary, our study finds that in the considered mass range of 10-365 GeV, the FCC-ee will increase the exclusion in the coupling $g'$ by one to two orders of magnitude. These results motivate further studies of leptophilic models for all the planned FCC-ee runs\footnote{Leptophilic scalars have only been considered at the $Z-$pole~\cite{Cesarotti:2024rbh}.}, constituting an important pillar for the physics case of the FCC-ee.

\section*{Acknowledgements}
We would like to thank Giacomo Polesello and Filip Zarnecki for useful discussions. 
The Swedish Research Council supports RGS (VR 2023-03403). BP and JZ are supported by the {\it Generalitat Valenciana} (Spain) through the {\it plan GenT} program (CIDEGENT/2019/068), by the Spanish Government (Agencia Estatal de
Investigaci\'on), ERDF funds from European Commission (MCIN/AEI/10.13039/501100011033, Grant No. PID2020-114473GB-I00), and by the Spanish Research Agency (Agencia Estatal
de Investigaci\'on, MCIU/AEI) through the grant IFIC Centro de Excelencia Severo Ochoa No. CEX2023-001292-S. RGS, BP and JZ have received support from the Consejo Superior de Investigaciones Cient\'ificas (CSIC) through de ``Red de Internacionalizaci\'on'' ILINK, grant No. 22043. BP thanks Laboratoire de Physique de
Clermont Auvergne for their hospitality during the final stages of this work.

The Feynman diagrams shown in this article have been drawn by means
of FeynGame~\cite{Harlander:2020cyh}.

\appendix
\section{Kinetic Mixing}
\label{app.kinmix}
While in these $L_x - L_y$ models discrete exchange symmetries $x \leftrightarrow y$ can prevent tree level mixing between the $Z'$ and the $Z, \gamma$ (or more appropriately, the $U(1)_Y$ B boson), the kinetic mixing would be induced by lepton loops, due to the lepton masses. So, we will assume that the kinetic term of our Lagrangian reads
\begin{equation}
{\cal L}_{kin} = \frac{1}{4} \Bigl [\hat{Z'_{\mu \nu}} \hat{Z'^{\mu \nu}} + \hat{B_{\mu \nu}} \hat{B^{\mu \nu}} + 2 \epsilon \hat{Z'_{\mu \nu}} \hat{B^{\mu \nu}}] \, ,
\end{equation}
where we followed the conventions of reference~\cite{Araki:2017wyg}. The expression for the 1-loop induced kinetic mixing is 
\begin{equation}
\epsilon = \Pi (q^2) = \frac{8 e g'}{(4 \pi)^2} \int_0^1 dx \,  x (1-x) \log \Biggl[ \frac{a- x (1-x)}{b - x (1-x)} \Biggr] =  \frac{8 e g'}{(4 \pi)^2}  I(a,b) \, ,
\end{equation}
where we have defined $a = m_{x}^2 / q^2, b = m_{y}^2 / q^2$. Since in our case $a, b \lesssim 1$ at first order in $a,b$ that $I(a,b) = b - a$. Hence $\epsilon$ is has a double suppression with respect to $g'$, first through the loop factor, and furthermore by the fact that the loop function is suppressed by the $a,b$ factors. For our range of masses, in the extreme case $\mzp = 10$ GeV we would obtain $\epsilon \sim 5 \times 10^{-4} \gp$. While this estimation enables us to ignore the kinetic mixing along this work, we will nonetheless describe here the impact on the neutral vector boson ($\gamma, Z, Z'$) couplings to fermions. These effects could be relevant if a non-zero tree-level kinetic mixing is present, which is beyond the scope of this work.

First and foremost, we need to obtain a kinetic term that is canonically normalized, namely perform a transformation
\begin{equation}
\hat{B}_{\mu \nu}  = B_{\mu \nu} + \alpha Z'_{\mu \nu}, \, \qquad \, 
\hat{Z'}_{\mu \nu} = \beta B_{\mu \nu} + \gamma Z'_{\mu \nu} \, ,
\end{equation}
such that the kinetic term would read
\begin{equation}
{\cal L}_{kin} = \frac{1}{4} \Bigl [Z'_{\mu \nu} Z'^{\mu \nu} + B_{\mu \nu} B^{\mu \nu} ] \, .
\end{equation}
It is trivial to find out that this can be achieved by choosing
\begin{equation}
\alpha = -\epsilon  (1-\epsilon^2)^{-1/2} \, , \qquad 
\beta=0\, , \qquad 
\gamma =  (1-\epsilon^2)^{-1/2} \, .
\end{equation}
Inserting this transformation into the covariant derivative, we obtain
\begin{equation}
D_{\mu} = \partial_{\mu} - i \Biggl\{ g_1 Y B_{\mu} + g_2 T_i W_{\mu}^{i} + \Bigl(\frac{g'(L_x - L_y)}{\sqrt{1-\epsilon^2}} - \frac{\epsilon g_1 Y}{\sqrt{1-\epsilon^2}} \Bigr) Z'_{\mu} \Biggr\} \, .
\end{equation}
The charged gauge bosons part proceeds as in the SM. We could diagonalize the 3x3 matrix from the neutral gauge bosons obtained from the covariant derivative terms, but it is more pedagogical to first perform the Weinberg rotation ($B_\mu = c_W \tilde{A}_\mu - s_W \tilde{Z}_{\mu}$, $Z_\mu = s_W \tilde{A}_\mu + c_W \tilde{Z}_{\mu}$, with $s_W = g_1 / \sqrt{g_1^2 +g_2^2}$ being the weak-mixing angle) to then obtain
\begin{equation}
\begin{split}
D_{\mu} - \partial_{\mu} = -i \Biggl\{ &\Bigl( g_1 c_W Y + g_2 s_W T_3 \Bigr) \tilde{A}_{\mu} + \Bigl( - g_1 s_W Y + g_2 c_W T_3 \Bigr) \tilde{Z}_{\mu}  \\
 &+ \Bigl(\frac{g'(L_x - L_y)}{\sqrt{1-\epsilon^2}} - \frac{\epsilon g_1 Y}{\sqrt{1-\epsilon^2}} \Bigr) \tilde{Z}'_{\mu} \Biggr\} \, ,
\end{split}
\end{equation}
where for convenience $Z'_{\mu}$ was renamed to $\tilde{Z}'_{\mu}$. Recalling that the SM Higgs is an $SU(2)_L$ doublet with $Y=1/2$, $Q=T_3 + Y$, that the Z' takes its mass from a new scalar $\phi$ with charge $L_x - L_y =2$ and vacuum expectation value $v_{\phi}$ (taking this convention from~\cite{Dasgupta:2023zrh}) and uncharged under the SM gauge groups ($T_3 \phi =0 = Y \phi)$ we would then obtain for the gauge boson masses 
\begin{equation}
    \frac{1}{2}\begin{pmatrix}
        \tilde{A}_\mu & \tilde{Z}_\mu & \tilde{Z}'_\mu
    \end{pmatrix}  \begin{pmatrix}
        0 & 0 & 0\\
        0 & M_{Z_0}^2 & \frac{\epsilon}{\sqrt{1-\epsilon^2}} s_W M_{Z_0}^2\\
        0 & \frac{\epsilon}{\sqrt{1-\epsilon^2}} s_W M_{Z_0}^2 & \frac{\epsilon^2}{1-\epsilon^2} s_W^2 M_{Z_0}^2+M_{Z'_0}^2
    \end{pmatrix}
    \begin{pmatrix}
        \tilde{A}^\mu\\
        \tilde{Z}^\mu\\
        \tilde{Z}'^\mu
    \end{pmatrix}, 
    \label{eq:LagNeutMass}
\end{equation}
where we have defined the tree level masses $M_{Z_0}^2=(g_1^2+g_2^2) v^2 /2$ and  $M_{Z'_0}=  2 g' v_{\phi} / \sqrt{1-\epsilon^2} $.
An additional rotation in angle $\alpha$ is needed to diagonalize the $Z-Z'$ mixing, but note that there is no mixture with the SM photon, this is because $U(1)_{em}$ remains unbroken. We can then write this rotation as
\begin{equation}
    \begin{pmatrix}
        \tilde{A}^\mu\\
        \tilde{Z}^\mu\\
        \tilde{Z}'^\mu
    \end{pmatrix}
    = \begin{pmatrix}
        1 & 0 & 0 \\
        0 & c_\alpha & s_{\alpha} \\
        0 & -s_\alpha & c_{\alpha} \\
    \end{pmatrix} 
    \begin{pmatrix}
        A^\mu\\
        Z^\mu\\
        Z'^\mu
    \end{pmatrix} ,
    \label{eq:LagNeutMass2}
\end{equation}
where $\alpha$ is given by
\begin{equation}
\tan 2 \alpha = \frac{2 \epsilon \sqrt{1- \epsilon^2} s_W M_{Z_0}^2}{M_{Z_0}^2 (1 - \epsilon^2 (1+s_W^2)) - M_{Z'_0}^2 (1- \epsilon^2)} \, ,
\end{equation}
which for small angles (expanding to first order in $\epsilon$) we obtain $\alpha = \epsilon s_W  / (1  - \Delta_Z^2) $  (where $\Delta_Z^2 = M_{Z'_0}^2 / M_{Z_0}^2$). Note that $\alpha$ is small unless the two gauge bosons are mass degenerate. With these final rotations the covariant derivative reads
\begin{equation}
D_{\mu} - \partial_{\mu} = -i \Biggl\{e Q A_{\mu} + \frac{g_2}{c_W} \Bigl[ T_3 - s_W^2 Q \Bigr] \tilde{Z}_{\mu} 
 + \Bigl[\frac{g'(L_x - L_y)}{\sqrt{1-\epsilon^2}} + \frac{g_2}{c_W} (\frac{\epsilon s_W (T_3 - Q)}{\sqrt{1-\epsilon^2}}) \Bigr] \tilde{Z}'_{\mu} \Biggr\} \, 
\end{equation}
after the Weinberg rotation, and 
\begin{equation}
\begin{split}
D_{\mu} - \partial_{\mu} = -i \Biggl\{& e Q A_{\mu} + \frac{g_2 c_{\alpha}}{c_W} \Bigl[ (1  - \frac{ t_{\alpha} \epsilon s_W}{\sqrt{1-\epsilon^2}} ) T_3 - (1 -  \frac{t_{\alpha} \epsilon s_W}{s_W^2 \sqrt{1-\epsilon^2}}) s_W^2 Q \\ & -  \frac{g' t_{\alpha} c_W}{g_2 \sqrt{1-\epsilon^2}}) (L_x - L_y) \Bigr] Z_{\mu} + 
 g^{\prime} \Bigl[ (\frac{g_2}{\gp c_W} (s_{\alpha} + c_{\alpha} \frac{\epsilon s_W}{\sqrt{1-\epsilon^2}} ) T_3  \\ & - (\frac{g_2 s_W^2}{\gp c_W} (s_{\alpha} + c_{\alpha} \frac{\epsilon s_W}{s_W^2 \sqrt{1-\epsilon^2}} )  Q  + (\frac{c_\alpha}{\sqrt{1-\epsilon^2}})  (L_x - L_y) \Bigr] Z'_{\mu} \Biggr\} \,  ,
 \end{split}
\end{equation}
after the final $\alpha$ rotation.

To obtain an insight on these expressions, we can work on the limit with small $\alpha$, working at first order in $\epsilon$. This works provided that $\Delta_Z$ does not gets close to one as to spoil the fact that $\alpha$ is small. In that case, $c_{\alpha} = 1$, $s_{\alpha} = \alpha = \epsilon s_W / (1-\Delta_Z^2)$ and hence we would find at this order
\begin{equation}
\begin{split}
D_{\mu} - \partial_{\mu} &= -i \Biggl\{ e Q A_{\mu} + \frac{g_2}{c_W} \Bigl[T_3 -  s_W^2 Q  -  \frac{\epsilon g' s_W c_W}{g_2 \sqrt{1-\epsilon^2} (1-\Delta_Z^2)}  (L_x - L_y) \Bigr] Z_{\mu} \\ &+ 
 g^{\prime} \Bigl[ \frac{\epsilon g_2 s_W \Delta_Z^2}{\gp c_W (1-\Delta_Z^2)} T_3 - \frac{\epsilon g_2 s_W}{\gp c_W } (1+ \frac{s_W^2}{1-\Delta_Z^2}   )  Q  +  (L_x - L_y) \Bigr] Z'_{\mu} \Biggr\} \,  .
 \end{split}
\end{equation}
Hence we see here that the $Z$ couplings to $T_3$ and $Q$ are unaffected at this order, while the coefficient accompanying lepton number goes as $\epsilon g'$ which is suppressed enough to be ignored. For the $Z'$ couplings, we see that both $T_3$ and $Q$ terms have a prefactor of $\epsilon/ \gp$, hence we can ignore them with respect to the coupling through lepton number\footnote{The induced coupling of the $Z'$ to quarks through mixing can also be ignored.}, as we assumed in the main text. 

\bibliographystyle{JHEP} 
\bibliography{ZLaLbFCCee}

\providecommand{\href}[2]{#2}\begingroup\raggedright\begin{thebibliography}{10}

\bibitem{FCC:2018evy}
{\bf FCC} Collaboration, A.~Abada et~al., {\it {FCC-ee: The Lepton Collider}: {Future Circular Collider Conceptual Design Report Volume 2}},  {\em Eur. Phys. J. ST} {\bf 228} (2019), no.~2 261--623.

\bibitem{ILC:2013jhg}
{\bf ILC} Collaboration, {\it {The International Linear Collider Technical Design Report - Volume 2: Physics}},  \href{http://arxiv.org/abs/1306.6352}{{\tt arXiv:1306.6352}}.

\bibitem{CEPCStudyGroup:2018ghi}
{\bf CEPC Study Group} Collaboration, M.~Dong et~al., {\it {CEPC Conceptual Design Report: Volume 2 - Physics \& Detector}},  \href{http://arxiv.org/abs/1811.10545}{{\tt arXiv:1811.10545}}.

\bibitem{CLICdp:2018cto}
{\bf CLICdp, CLIC} Collaboration, T.~K. Charles et~al., {\it {The Compact Linear Collider (CLIC) - 2018 Summary Report}},  \href{http://arxiv.org/abs/1812.06018}{{\tt arXiv:1812.06018}}.

\bibitem{MuonCollider:2022xlm}
{\bf Muon Collider} Collaboration, J.~de~Blas et~al., {\it {The physics case of a 3 TeV muon collider stage}},  \href{http://arxiv.org/abs/2203.07261}{{\tt arXiv:2203.07261}}.

\bibitem{Mohapatra:1980de}
R.~N. Mohapatra and R.~E. Marshak, {\it {PHENOMENOLOGY OF NEUTRON OSCILLATIONS}},  {\em Phys. Lett. B} {\bf 94} (1980) 183. [Erratum: Phys.Lett.B 96, 444--444 (1980)].

\bibitem{Foot:1990mn}
R.~Foot, {\it {New Physics From Electric Charge Quantization?}},  {\em Mod. Phys. Lett. A} {\bf 6} (1991) 527--530.

\bibitem{He:1990pn}
X.~G. He, G.~C. Joshi, H.~Lew, and R.~R. Volkas, {\it {NEW Z-prime PHENOMENOLOGY}},  {\em Phys. Rev. D} {\bf 43} (1991) 22--24.

\bibitem{Foot:1994vd}
R.~Foot, X.~G. He, H.~Lew, and R.~R. Volkas, {\it {Model for a light Z-prime boson}},  {\em Phys. Rev. D} {\bf 50} (1994) 4571--4580, [\href{http://arxiv.org/abs/hep-ph/9401250}{{\tt hep-ph/9401250}}].

\bibitem{Biswas:2016yan}
A.~Biswas, S.~Choubey, and S.~Khan, {\it {Neutrino Mass, Dark Matter and Anomalous Magnetic Moment of Muon in a $U(1)_{L_{\mu}-L_{\tau}}$ Model}},  {\em JHEP} {\bf 09} (2016) 147, [\href{http://arxiv.org/abs/1608.04194}{{\tt arXiv:1608.04194}}].

\bibitem{Baek:2015mna}
S.~Baek, H.~Okada, and K.~Yagyu, {\it {Flavour Dependent Gauged Radiative Neutrino Mass Model}},  {\em JHEP} {\bf 04} (2015) 049, [\href{http://arxiv.org/abs/1501.01530}{{\tt arXiv:1501.01530}}].

\bibitem{Heeck:2011wj}
J.~Heeck and W.~Rodejohann, {\it {Gauged $L_\mu - L_\tau$ Symmetry at the Electroweak Scale}},  {\em Phys. Rev. D} {\bf 84} (2011) 075007, [\href{http://arxiv.org/abs/1107.5238}{{\tt arXiv:1107.5238}}].

\bibitem{Patra:2016shz}
S.~Patra, S.~Rao, N.~Sahoo, and N.~Sahu, {\it {Gauged $U(1)_{L_\mu - L_\tau}$ model in light of muon $g-2$ anomaly, neutrino mass and dark matter phenomenology}},  {\em Nucl. Phys. B} {\bf 917} (2017) 317--336, [\href{http://arxiv.org/abs/1607.04046}{{\tt arXiv:1607.04046}}].

\bibitem{Arcadi:2018tly}
G.~Arcadi, T.~Hugle, and F.~S. Queiroz, {\it {The Dark $L_\mu - L_\tau$ Rises via Kinetic Mixing}},  {\em Phys. Lett. B} {\bf 784} (2018) 151--158, [\href{http://arxiv.org/abs/1803.05723}{{\tt arXiv:1803.05723}}].

\bibitem{Altmannshofer:2016jzy}
W.~Altmannshofer, S.~Gori, S.~Profumo, and F.~S. Queiroz, {\it {Explaining dark matter and B decay anomalies with an $L_\mu - L_\tau$ model}},  {\em JHEP} {\bf 12} (2016) 106, [\href{http://arxiv.org/abs/1609.04026}{{\tt arXiv:1609.04026}}].

\bibitem{Gninenko:2001hx}
S.~N. Gninenko and N.~V. Krasnikov, {\it {The Muon anomalous magnetic moment and a new light gauge boson}},  {\em Phys. Lett. B} {\bf 513} (2001) 119, [\href{http://arxiv.org/abs/hep-ph/0102222}{{\tt hep-ph/0102222}}].

\bibitem{Ma:2001md}
E.~Ma, D.~P. Roy, and S.~Roy, {\it {Gauged L(mu) - L(tau) with large muon anomalous magnetic moment and the bimaximal mixing of neutrinos}},  {\em Phys. Lett. B} {\bf 525} (2002) 101--106, [\href{http://arxiv.org/abs/hep-ph/0110146}{{\tt hep-ph/0110146}}].

\bibitem{Baek:2001kca}
S.~Baek, N.~G. Deshpande, X.~G. He, and P.~Ko, {\it {Muon anomalous g-2 and gauged L(muon) - L(tau) models}},  {\em Phys. Rev. D} {\bf 64} (2001) 055006, [\href{http://arxiv.org/abs/hep-ph/0104141}{{\tt hep-ph/0104141}}].

\bibitem{Dasgupta:2023zrh}
A.~Dasgupta, P.~S.~B. Dev, T.~Han, R.~Padhan, S.~Wang, and K.~Xie, {\it {Searching for heavy leptophilic Z': from lepton colliders to gravitational waves}},  {\em JHEP} {\bf 12} (2023) 011, [\href{http://arxiv.org/abs/2308.12804}{{\tt arXiv:2308.12804}}].

\bibitem{Barik:2024kwv}
A.~K. Barik, S.~K. Rai, and A.~Srivastava, {\it {Discovering an invisible Z' at the muon collider}},  \href{http://arxiv.org/abs/2408.14396}{{\tt arXiv:2408.14396}}.

\bibitem{Yue:2024kwo}
C.-X. Yue, Y.~Li, M.~Wang, and X.~Zhang, {\it {Searching for the light leptophilic gauge boson $Z_x$ via four-lepton final states at the CEPC}},  \href{http://arxiv.org/abs/2402.00619}{{\tt arXiv:2402.00619}}.

\bibitem{Goudelis:2023yni}
A.~Goudelis, J.~Kriewald, E.~Pinsard, and A.~M. Teixeira, {\it {cLFV leptophilic $Z^\prime$ as a dark matter portal: prospects for colliders}},  \href{http://arxiv.org/abs/2312.14103}{{\tt arXiv:2312.14103}}.

\bibitem{Hapitas:2021ilr}
T.~Hapitas, D.~Tuckler, and Y.~Zhang, {\it {General kinetic mixing in gauged U(1)L\ensuremath{\mu}-L\ensuremath{\tau} model for muon g-2 and dark matter}},  {\em Phys. Rev. D} {\bf 105} (2022), no.~1 016014, [\href{http://arxiv.org/abs/2108.12440}{{\tt arXiv:2108.12440}}].

\bibitem{DeRomeri:2024dbv}
V.~De~Romeri, D.~K. Papoulias, and C.~A. Ternes, {\it {Light vector mediators at direct detection experiments}},  \href{http://arxiv.org/abs/2402.05506}{{\tt arXiv:2402.05506}}.

\bibitem{Lees:2014xha}
{\bf BaBar} Collaboration, J.~P. Lees et~al., {\it {Search for a Dark Photon in $e^+e^-$ Collisions at BaBar}},  {\em Phys. Rev. Lett.} {\bf 113} (2014), no.~20 201801, [\href{http://arxiv.org/abs/1406.2980}{{\tt arXiv:1406.2980}}].

\bibitem{Ilten:2018crw}
P.~Ilten, Y.~Soreq, M.~Williams, and W.~Xue, {\it {Serendipity in dark photon searches}},  {\em JHEP} {\bf 06} (2018) 004, [\href{http://arxiv.org/abs/1801.04847}{{\tt arXiv:1801.04847}}].

\bibitem{CMS:2018yxg}
{\bf CMS} Collaboration, A.~M. Sirunyan et~al., {\it {Search for an $L_{\mu}-L_{\tau}$ gauge boson using Z$\to4\mu$ events in proton-proton collisions at $\sqrt{s} =$ 13 TeV}},  {\em Phys. Lett. B} {\bf 792} (2019) 345--368, [\href{http://arxiv.org/abs/1808.03684}{{\tt arXiv:1808.03684}}].

\bibitem{ATLAS:2023vxg}
{\bf ATLAS} Collaboration, G.~Aad et~al., {\it {Search for a new Z' gauge boson in $4\mu$ events with the ATLAS experiment}},  {\em JHEP} {\bf 07} (2023) 090, [\href{http://arxiv.org/abs/2301.09342}{{\tt arXiv:2301.09342}}].

\bibitem{ATLAS:2024uvu}
{\bf ATLAS} Collaboration, G.~Aad et~al., {\it {Search for a new $Z'$ gauge boson via the $pp \rightarrow W^{\pm(*)} \rightarrow Z' \mu^{\pm} \nu \rightarrow \mu^{\pm}\mu^{\mp}\mu^{\pm}\nu$ process in $pp$ collisions at $\sqrt{s}=13$ TeV with the ATLAS detector}},  \href{http://arxiv.org/abs/2402.15212}{{\tt arXiv:2402.15212}}.

\bibitem{Bauer:2018onh}
M.~Bauer, P.~Foldenauer, and J.~Jaeckel, {\it {Hunting All the Hidden Photons}},  {\em JHEP} {\bf 07} (2018) 094, [\href{http://arxiv.org/abs/1803.05466}{{\tt arXiv:1803.05466}}].

\bibitem{IceCube:2022pbe}
{\bf IceCube} Collaboration, R.~Abbasi et~al., {\it {Non-standard neutrino interactions in IceCube}},  {\em PoS} {\bf EPS-HEP2021} (2022) 245.

\bibitem{Fox:2011fx}
P.~J. Fox, R.~Harnik, J.~Kopp, and Y.~Tsai, {\it {LEP Shines Light on Dark Matter}},  {\em Phys. Rev. D} {\bf 84} (2011) 014028, [\href{http://arxiv.org/abs/1103.0240}{{\tt arXiv:1103.0240}}].

\bibitem{DELPHI:2003dlq}
{\bf DELPHI} Collaboration, J.~Abdallah et~al., {\it {Photon events with missing energy in e+ e- collisions at s**(1/2) = 130-GeV to 209-GeV}},  {\em Eur. Phys. J. C} {\bf 38} (2005) 395--411, [\href{http://arxiv.org/abs/hep-ex/0406019}{{\tt hep-ex/0406019}}].

\bibitem{DELPHI:2008uka}
{\bf DELPHI} Collaboration, J.~Abdallah et~al., {\it {Search for one large extra dimension with the DELPHI detector at LEP}},  {\em Eur. Phys. J. C} {\bf 60} (2009) 17--23, [\href{http://arxiv.org/abs/0901.4486}{{\tt arXiv:0901.4486}}].

\bibitem{Buckley:2011vc}
M.~R. Buckley, D.~Hooper, J.~Kopp, and E.~Neil, {\it {Light Z' Bosons at the Tevatron}},  {\em Phys. Rev. D} {\bf 83} (2011) 115013, [\href{http://arxiv.org/abs/1103.6035}{{\tt arXiv:1103.6035}}].

\bibitem{Kalinowski:2020lhp}
J.~Kalinowski, W.~Kotlarski, P.~Sopicki, and A.~F. Zarnecki, {\it {Simulating hard photon production with WHIZARD}},  {\em Eur. Phys. J. C} {\bf 80} (2020), no.~7 634, [\href{http://arxiv.org/abs/2004.14486}{{\tt arXiv:2004.14486}}].

\bibitem{Kalinowski:2021tyr}
J.~Kalinowski, W.~Kotlarski, K.~Mekala, P.~Sopicki, and A.~F. Zarnecki, {\it {Sensitivity of future linear $\hbox {e}^+\hbox {e}^-$ colliders to processes of dark matter production with light mediator exchange}},  {\em Eur. Phys. J. C} {\bf 81} (2021), no.~10 955, [\href{http://arxiv.org/abs/2107.11194}{{\tt arXiv:2107.11194}}].

\bibitem{limitsILC}
A.~F. Zarnecki, {\it Light mediator searches with mono-photon signature},  2022.

\bibitem{Manohar:2017eqh}
A.~V. Manohar, P.~Nason, G.~P. Salam, and G.~Zanderighi, {\it {The Photon Content of the Proton}},  {\em JHEP} {\bf 12} (2017) 046, [\href{http://arxiv.org/abs/1708.01256}{{\tt arXiv:1708.01256}}].

\bibitem{Buonocore:2020nai}
L.~Buonocore, P.~Nason, F.~Tramontano, and G.~Zanderighi, {\it {Leptons in the proton}},  {\em JHEP} {\bf 08} (2020), no.~08 019, [\href{http://arxiv.org/abs/2005.06477}{{\tt arXiv:2005.06477}}].

\bibitem{Alwall:2014hca}
J.~Alwall, R.~Frederix, S.~Frixione, V.~Hirschi, F.~Maltoni, O.~Mattelaer, H.~S. Shao, T.~Stelzer, P.~Torrielli, and M.~Zaro, {\it {The automated computation of tree-level and next-to-leading order differential cross sections, and their matching to parton shower simulations}},  {\em JHEP} {\bf 07} (2014) 079, [\href{http://arxiv.org/abs/1405.0301}{{\tt arXiv:1405.0301}}].

\bibitem{Agapov:2022bhm}
I.~Agapov et~al., {\it {Future Circular Lepton Collider FCC-ee: Overview and Status}},  in {\em {Snowmass 2021}}, 3, 2022.
\newblock \href{http://arxiv.org/abs/2203.08310}{{\tt arXiv:2203.08310}}.

\bibitem{Bernardi:2022hny}
G.~Bernardi et~al., {\it {The Future Circular Collider: a Summary for the US 2021 Snowmass Process}},  \href{http://arxiv.org/abs/2203.06520}{{\tt arXiv:2203.06520}}.

\bibitem{FCCMidTerm}
B.~Auchmann, W.~Bartmann, M.~Benedikt, J.-P. Burnet, P.~Craievich, M.~Giovannozzi, C.~Grojean, J.~Gutleber, K.~Hanke, P.~Janot, M.~Mangano, J.~Osborne, J.~Poole, T.~Raubenheimer, T.~Watson, F.~Zimmermann, R.~Losito, et~al., {\it {FCC Midterm Report}},  2024.

\bibitem{Bierlich:2022pfr}
C.~Bierlich et~al., {\it {A comprehensive guide to the physics and usage of PYTHIA 8.3}},  {\em SciPost Phys. Codeb.} {\bf 2022} (2022) 8, [\href{http://arxiv.org/abs/2203.11601}{{\tt arXiv:2203.11601}}].

\bibitem{deFavereau:2013fsa}
{\bf DELPHES 3} Collaboration, J.~de~Favereau, C.~Delaere, P.~Demin, A.~Giammanco, V.~Lema\^\i{}tre, A.~Mertens, and M.~Selvaggi, {\it {DELPHES 3, A modular framework for fast simulation of a generic collider experiment}},  {\em JHEP} {\bf 02} (2014) 057, [\href{http://arxiv.org/abs/1307.6346}{{\tt arXiv:1307.6346}}].

\bibitem{Conte:2012fm}
E.~Conte, B.~Fuks, and G.~Serret, {\it {MadAnalysis 5, A User-Friendly Framework for Collider Phenomenology}},  {\em Comput. Phys. Commun.} {\bf 184} (2013) 222--256, [\href{http://arxiv.org/abs/1206.1599}{{\tt arXiv:1206.1599}}].

\bibitem{Conte:2014zja}
E.~Conte, B.~Dumont, B.~Fuks, and C.~Wymant, {\it {Designing and recasting LHC analyses with MadAnalysis 5}},  {\em Eur. Phys. J. C} {\bf 74} (2014), no.~10 3103, [\href{http://arxiv.org/abs/1405.3982}{{\tt arXiv:1405.3982}}].

\bibitem{Dumont:2014tja}
B.~Dumont, B.~Fuks, S.~Kraml, S.~Bein, G.~Chalons, E.~Conte, S.~Kulkarni, D.~Sengupta, and C.~Wymant, {\it {Toward a public analysis database for LHC new physics searches using MADANALYSIS 5}},  {\em Eur. Phys. J. C} {\bf 75} (2015), no.~2 56, [\href{http://arxiv.org/abs/1407.3278}{{\tt arXiv:1407.3278}}].

\bibitem{Conte:2018vmg}
E.~Conte and B.~Fuks, {\it {Confronting new physics theories to LHC data with MADANALYSIS 5}},  {\em Int. J. Mod. Phys. A} {\bf 33} (2018), no.~28 1830027, [\href{http://arxiv.org/abs/1808.00480}{{\tt arXiv:1808.00480}}].

\bibitem{Araz:2019otb}
J.~Y. Araz, M.~Frank, and B.~Fuks, {\it {Reinterpreting the results of the LHC with MadAnalysis 5: uncertainties and higher-luminosity estimates}},  {\em Eur. Phys. J. C} {\bf 80} (2020), no.~6 531, [\href{http://arxiv.org/abs/1910.11418}{{\tt arXiv:1910.11418}}].

\bibitem{Araz:2020lnp}
J.~Y. Araz, B.~Fuks, and G.~Polykratis, {\it {Simplified fast detector simulation in MADANALYSIS 5}},  {\em Eur. Phys. J. C} {\bf 81} (2021), no.~4 329, [\href{http://arxiv.org/abs/2006.09387}{{\tt arXiv:2006.09387}}].

\bibitem{Backovic:2015soa}
M.~Backovi\'c, M.~Kr\"amer, F.~Maltoni, A.~Martini, K.~Mawatari, and M.~Pellen, {\it {Higher-order QCD predictions for dark matter production at the LHC in simplified models with s-channel mediators}},  {\em Eur. Phys. J. C} {\bf 75} (2015), no.~10 482, [\href{http://arxiv.org/abs/1508.05327}{{\tt arXiv:1508.05327}}].

\bibitem{Albert:2017onk}
A.~Albert et~al., {\it {Recommendations of the LHC Dark Matter Working Group: Comparing LHC searches for dark matter mediators in visible and invisible decay channels and calculations of the thermal relic density}},  {\em Phys. Dark Univ.} {\bf 26} (2019) 100377, [\href{http://arxiv.org/abs/1703.05703}{{\tt arXiv:1703.05703}}].

\bibitem{Cesarotti:2024rbh}
C.~Cesarotti and G.~Krnjaic, {\it {Hitting the Thermal Target for Leptophilic Dark Matter}},  \href{http://arxiv.org/abs/2404.02906}{{\tt arXiv:2404.02906}}.

\bibitem{Harlander:2020cyh}
R.~V. Harlander, S.~Y. Klein, and M.~Lipp, {\it {FeynGame}},  {\em Comput. Phys. Commun.} {\bf 256} (2020) 107465, [\href{http://arxiv.org/abs/2003.00896}{{\tt arXiv:2003.00896}}].

\bibitem{Araki:2017wyg}
T.~Araki, S.~Hoshino, T.~Ota, J.~Sato, and T.~Shimomura, {\it {Detecting the $L_{\mu}-L_{\tau}$ gauge boson at Belle II}},  {\em Phys. Rev. D} {\bf 95} (2017), no.~5 055006, [\href{http://arxiv.org/abs/1702.01497}{{\tt arXiv:1702.01497}}].

\end{thebibliography}\endgroup

\end{document}